\documentclass[12pt,a4paper]{report}

\usepackage[utf8]{inputenc}  % MOVE THIS BEFORE csquotes
\usepackage[T1]{fontenc}
\usepackage[english]{babel}
\usepackage{csquotes}  % Now it comes after inputenc
\usepackage{textcomp}
\usepackage{siunitx} % For degree symbol
\usepackage{textgreek} % For Greek letters% For Greek letters

\usepackage{geometry}
\usepackage{setspace}
\usepackage{newtxtext,newtxmath}
\usepackage{booktabs}
\usepackage{graphicx}
\usepackage{float}
\usepackage{caption}

\usepackage[hidelinks]{hyperref}

\usepackage{fancyhdr}
\usepackage{subcaption}
\usepackage{enumitem}
\setlist[itemize]{label=•, left=0pt, itemsep=0.7em}

\usepackage[backend=biber,style=numeric,sorting=none]{biblatex}
\newcommand{\authors}{Mohammed El Abdioui}
\newcommand{\supervisors}{Rachid Ilmen}
\newcommand{\location}{Casablanca - Morocco}
\newcommand{\reportDate}{October 2025}

\begin{document}

% ====== Title Page ======
\begin{titlepage}
    \centering
    \vspace*{1cm}

    % ====== EHTP Logo ======
    \includegraphics[width=0.25\textwidth]{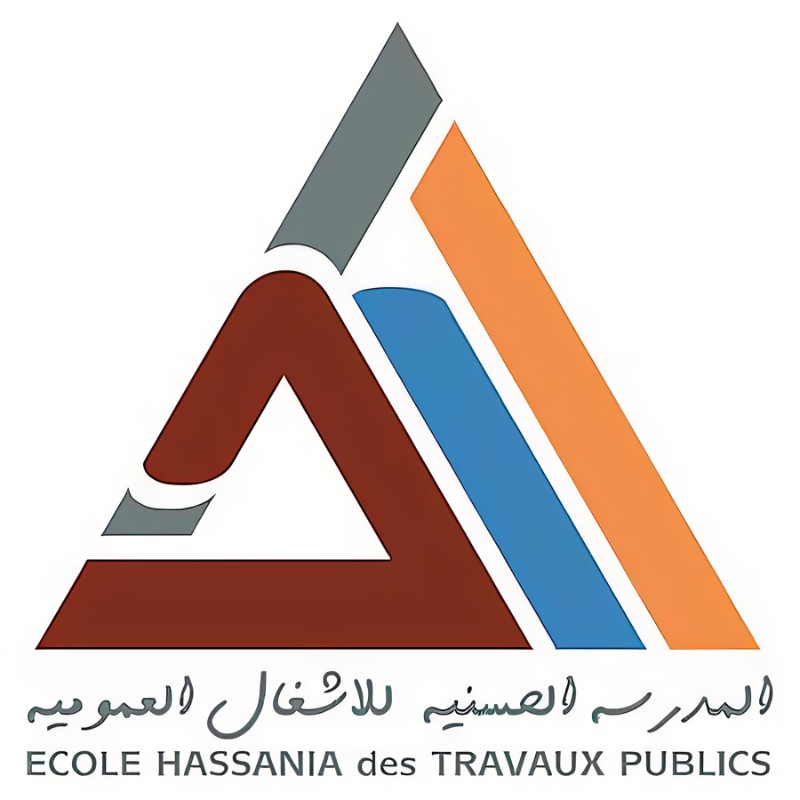}\\[1cm]

    % ====== School ======
    {\small  The Hassania School of Public Works }\\[0.5cm]

    % ====== Title ======
    {\LARGE \textbf{Effects of Climate Change on Moroccan Coastal Upwelling:}\\[0.3cm]
    {\LARGE \textbf{Relationships between the NAO, Upwelling Index, and Sea Surface Temperature (1978-2024)}}\par}
    \vspace{2.5cm}

    % ====== Authors ======
    \textbf{Prepared by:}\\
    \authors\\[1.5cm]

    % ====== Supervisors ======
    \textbf{Supervised by:}\\
    \supervisors\\[3cm]

    % ====== Location and Date ======
    \vfill
    {\location}\\
    {\reportDate}
\end{titlepage}

\tableofcontents
\newpage

% Add List of Figures
\listoffigures
\newpage
% =====================================================

\chapter{General Introduction}

\section{Subject Context}

Climate change currently represents one of the greatest environmental challenges on a global scale. Its effects are particularly evident in coastal areas, where interactions between the atmosphere, ocean, and biosphere influence marine ecosystem dynamics.

Among the ocean processes most sensitive to these changes is \textbf{coastal upwelling}, a phenomenon where deep, cold, nutrient-rich waters rise to the surface, thus stimulating biological productivity and regulating regional climate.

The \textbf{Moroccan Atlantic coast}, stretching approximately 3500 km and representing one of the world's most dynamic upwelling zones, constitutes a privileged study area for assessing the impact of global warming on ocean-atmosphere processes. The intensity and variability of this upwelling directly influence fishery resources and, consequently, local economic activities such as:

\begin{itemize}  
    \item \textbf{Artisanal fishing}, characterized by small traditional boats (barques) and local practices, which heavily depends on fish availability near the coastline.  
    \item \textbf{Industrial fishing}, more industrial and destined for export or national markets, which exploits resources further out at sea and contributes significantly to the national economy.  
\end{itemize}

Today, the Moroccan climate is characterized by \textbf{structural drought}, which has induced a \textbf{shift in rainfall patterns} across the national territory, leading to \textbf{spatio-temporal variation in precipitation} across the country. These climatic conditions increase the sensitivity of coastal ecosystems and economic sectors dependent on marine resources.

In this context, the joint analysis of the \textbf{NAO}, the \textbf{Coastal Upwelling Index (CUI)}, and \textbf{Sea Surface Temperature (SST)} over a long period (1978-2024) allows for a better understanding of trends, climate variability mechanisms, and their potential impacts on coastal ecosystems and economic sectors related to fishing along the Moroccan coast. This study focuses specifically on the 3500 km of the Atlantic coastline, to provide a comprehensive view of the spatio-temporal variability of upwelling and its effects on artisanal and industrial fishing.

\begin{figure}[H]
    \centering
    \includegraphics[width=0.9\textwidth]{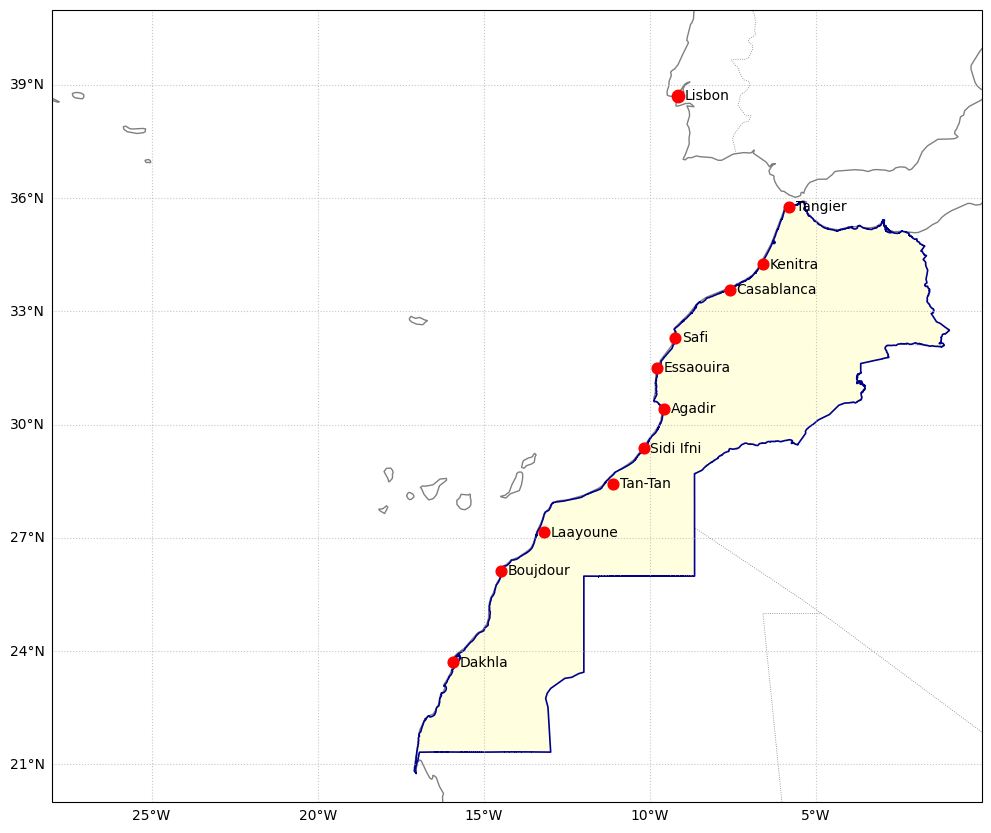} 
    \caption{Geographic representation of the Atlantic coast of Morocco, indicating the main coastal cities studied along the coastline, from Dakhla to Tangier. The coastlines of neighboring countries are included to provide regional context, with Lisbon (Portugal) used as a reference point.}
     \label{fig:morocco_map}
\end{figure}

\section{Problem Statement}

How do climate change and atmospheric variability, represented by the North Atlantic Oscillation (NAO), influence the dynamics of upwelling along the Moroccan Atlantic coast? And what are the potential consequences for sea surface temperature, marine productivity, coastal ecosystems, and dependent socio-economic activities such as fishing, aquaculture, and coastal tourism?

\section{General Objective}

This work aims to \textbf{analyze the impact of climate change and atmospheric variability on the dynamics of Moroccan coastal upwelling}, to better understand the interactions between the atmosphere and the ocean in the context of global warming.

\section{Specific Objectives}

More specifically, the objectives are to:
\begin{enumerate}
    \item Analyze the temporal evolution of CUI, SST, and NAO over the period 1978-2024;
    \item Study correlations between these variables to identify dominant trends and associated climate regimes;
    \item Assess the influence of climate change on the intensity and seasonality of Moroccan upwelling;
    \item Highlight the environmental and socio-economic implications of these evolutions, particularly on marine productivity and dependent sectors such as fishing and coastal tourism.
\end{enumerate}

\chapter{Materials and Methods}

\section{Scientific Instruments Used}

This study was conducted using the following software tools and programming environments:

\subsection{Languages and Environments}
\begin{itemize}
    \item \textbf{Python 3.9}: Main programming language for data analysis and scientific computations
    \item \textbf{Jupyter Notebook}: Interactive development environment for prototyping and exploratory analysis
\end{itemize}

\subsection{Python Scientific Libraries}
\begin{itemize}
    \item \textbf{xarray (v0.20.1)} and \textbf{netCDF4 (v1.6.2)}: Reading and manipulation of NetCDF files containing ERA5 data
    \item \textbf{NumPy (v1.23.0)} and \textbf{pandas (v1.5.0)}: Numerical computations and tabular data processing
    \item \textbf{SciPy (v1.9.0)}: Statistical analyses and advanced scientific functions
    \item \textbf{statsmodels (v0.13.2)}: Implementation of Granger causality tests and other temporal analyses
    \item \textbf{Matplotlib (v3.6.0)} and \textbf{Seaborn (v0.12.0)}: Data visualization and graph creation
\end{itemize}

\subsection{Data Used}

The data used in this study come from the \textbf{ERA5 reanalysis} produced by the \textit{European Centre for Medium-Range Weather Forecasts} (ECMWF). They were obtained via the Copernicus Climate Data Store (CDS) portal.

\begin{itemize}
    \item \textbf{Product type}: Monthly averaged reanalysis
    \item \textbf{Variables used}:
    \begin{itemize}
        \item 10m u-component of wind
        \item 10m v-component of wind
        \item Mean sea level pressure (SLP)
        \item Sea surface temperature (SST)
    \end{itemize}
    \item \textbf{Temporal period}: 1978-2024 (42 continuous years of data)
    \item \textbf{Temporal resolution}: Monthly data
    \item \textbf{Spatial resolution}: ERA5 grid at \SI{0.25}{\degree} $\times$ \SI{0.25}{\degree} (approximately 28 km), covering the entire North Atlantic, including the Moroccan Atlantic coast
    \item \textbf{File format}: Uncompressed \texttt{NetCDF4 (Experimental)}
\end{itemize}

These data offer essential spatio-temporal consistency for studying interactions between the atmosphere and the ocean. The selected variables allow examination of the physical mechanisms governing the \textbf{variability of Moroccan coastal upwelling}, in relation to the \textbf{North Atlantic Oscillation (NAO)} and \textbf{sea surface temperature (SST)} over the period 1978-2024, the main analysis period used in this work.

\section{Methodological Approach}

\subsection{Calculation of Climate Indices}

\subsubsection{NAO Index}
The NAO index was calculated according to the standard approach based on normalized pressure difference between the Azores (37°N-23°W) and Iceland (65°N-22°W):

\begin{enumerate}
    \item Extraction of SLP data at representative grid points
    \item Calculation of anomalies relative to the reference period 1981-2010
    \item Monthly normalization of anomalies to eliminate seasonal effects
    \item Calculation of normalized difference: 
    \begin{equation}
    NAO = \frac{P_{\text{Azores}} - \mu_{\text{Azores}}}{\sigma_{\text{Azores}}} - \frac{P_{\text{Iceland}} - \mu_{\text{Iceland}}}{\sigma_{\text{Iceland}}}
    \end{equation}
\end{enumerate}

\subsubsection{Upwelling Index}
The Moroccan coastal upwelling index was calculated according to the formulation by Hilmi et al. (2022):

\begin{align}
M_x & = \frac{\rho_{\text{air}} C_d |V| u_{10}}{\rho_w f} \\
M_y & = \frac{\rho_{\text{air}} C_d |V| v_{10}}{\rho_w f} \\
UI & = -M_y \times 100
\end{align}

where:
\begin{itemize}
    \item $\rho_{\text{air}} = \SI{1.22}{\kilo\gram\per\cubic\meter}$, $\rho_w = \SI{1025}{\kilo\gram\per\cubic\meter}$
    \item $C_d = 1.4 \times 10^{-3}$ (drag coefficient)
    \item $f = 2\omega \sin(\varphi)$ (Coriolis parameter)
    \item $u_{10}$, $v_{10}$: zonal and meridional components of wind at 10 m
\end{itemize}

\subsubsection{Ekman Transport}
The offshore Ekman transport ($Q_E$) was calculated to quantify water volume transported perpendicular to the coast:

\begin{align}
\tau_x &= \rho_{\text{air}} C_d u_{10} \sqrt{u_{10}^2 + v_{10}^2} \\
\tau_y &= \rho_{\text{air}} C_d v_{10} \sqrt{u_{10}^2 + v_{10}^2} \\
Q_{E_x} &= \frac{\tau_y}{\rho_w f} \\
Q_{E_y} &= -\frac{\tau_x}{\rho_w f} \\
|Q_E| &= \sqrt{Q_{E_x}^2 + Q_{E_y}^2}
\end{align}

where $\tau_x$, $\tau_y$ are wind stress components (N/m²). The upwelling index (UI) is equivalent to $-Q_{E_y} \times 100$.

\subsection{Statistical Analysis}

\subsubsection{Data Preprocessing}
\begin{itemize}
    \item Normalization of time series: 
    \begin{equation}
    X_{\text{norm}} = \frac{X - \mu}{\sigma}
    \end{equation}
    \item Seasonal aggregation according to meteorological seasons:
    \begin{itemize}
        \item Winter (DJF): December, January, February
        \item Spring (MAM): March, April, May
        \item Summer (JJA): June, July, August
        \item Autumn (SON): September, October, November
    \end{itemize}
\end{itemize}

\subsubsection{Granger Causality Analysis}

The Granger causality test procedure between the NAO index, upwelling, and SST follows the methodology established by Kaufmann and Stern (1997). This approach includes two main steps and allows identification of the presence and direction of causal relationships.

\paragraph{VAR Modeling}
In the first step, interactions between NAO, upwelling, and SST anomalies are described using a vector autoregressive (VAR) model. For each variable pair, the equation system is written:

\begin{align}
\text{NAO}_t &= \alpha_1 + \sum_{i=1}^s \beta_{1i} \text{NAO}_{t-i} + \sum_{i=1}^s \gamma_{1i} \text{SST}_{t-i} + \epsilon_{1t} \label{eq:nao_var} \\
\text{SST}_t &= \alpha_2 + \sum_{i=1}^s \beta_{2i} \text{NAO}_{t-i} + \sum_{i=1}^s \gamma_{2i} \text{SST}_{t-i} + \epsilon_{2t} \label{eq:sst_var}
\end{align}

Similarly, for relationships between upwelling and SST:

\begin{align}
\text{UI}_t &= \alpha_3 + \sum_{i=1}^s \beta_{3i} \text{UI}_{t-i} + \sum_{i=1}^s \gamma_{3i} \text{SST}_{t-i} + \epsilon_{3t} \label{eq:ui_var} \\
\text{SST}_t &= \alpha_4 + \sum_{i=1}^s \beta_{4i} \text{UI}_{t-i} + \sum_{i=1}^s \gamma_{4i} \text{SST}_{t-i} + \epsilon_{4t} \label{eq:sst_ui_var}
\end{align}

where $\alpha$, $\beta$, and $\gamma$ are regression coefficients, $\epsilon$ represents error terms, and $s$ is the lag length determined by a likelihood ratio test (Sims, 1980).

\paragraph{Seasonal Specification}
To account for seasonal variability in ocean-atmosphere interactions, we adapted the basic VAR equations for each meteorological season (DJF, MAM, JJA, SON). For example, for winter:

\begin{align}
\text{NAO}_{\text{Winter}_t} &= \alpha_1 + \sum_{i=1}^s \beta_{1i} \text{NAO}_{\text{Winter}_{t-i}} + \sum_{i=1}^s \gamma_{1i} \text{SST}_{\text{Winter}_{t-i}} + \epsilon_{1t} \label{eq:nao_winter} \\
\text{SST}_{\text{Winter}_t} &= \alpha_2 + \sum_{i=1}^s \beta_{2i} \text{NAO}_{\text{Winter}_{t-i}} + \sum_{i=1}^s \gamma_{2i} \text{SST}_{\text{Winter}_{t-i}} + \epsilon_{2t} \label{eq:sst_winter}
\end{align}

\paragraph{Causality Test}
To determine the direction of causality, we estimate restricted forms of the equations by eliminating the potential causal variable. For example, to test if SST causes NAO variability, we estimate the restricted equation:

\begin{equation}
\text{NAO}_t = \alpha'_1 + \sum_{i=1}^s \beta'_{1i} \text{NAO}_{t-i} + \epsilon'_{1t} \label{eq:nao_restricted}
\end{equation}

\paragraph{Test Statistic}
Statistical significance is evaluated using an F-statistic calculated as:

\begin{equation}
F = \frac{(RSS_r - RSS_u)/s}{RSS_u/(T - k)} \label{eq:f_stat}
\end{equation}

where:
\begin{itemize}
    \item $RSS_r$ and $RSS_u$ are residual sums of squares of restricted and unrestricted models
    \item $s$ is the number of coefficients restricted to zero
    \item $T$ is the number of observations
    \item $k$ is the total number of regressors in the unrestricted model
\end{itemize}

\paragraph{Complementary Causality Measure}
In addition to the F-statistic, we calculate the difference in coefficients of determination ($\Delta R^2$) to quantify the improvement in variance explanation:

\begin{equation}
\Delta R^2 = R^2_u - R^2_r = \frac{RSS_r - RSS_u}{RSS_r} \label{eq:delta_r2}
\end{equation}

where:
\begin{itemize}
    \item $R^2_u = 1 - \frac{RSS_u}{TSS}$ is the coefficient of determination of the unrestricted model
    \item $R^2_r = 1 - \frac{RSS_r}{TSS}$ is the coefficient of determination of the restricted model
    \item $TSS$ is the total sum of squares
\end{itemize}

This $\Delta R^2$ measure represents the additional proportion of variance explained by including lags of the causal variable in the model.

\paragraph{Interpretation}
A statistically significant F-value (threshold $p < 0.05$) rejects the null hypothesis of non-causality. This indicates that lagged values of the eliminated variable contain unique information for predicting the dependent variable, beyond that contained in its own past values.

The F-statistic evaluates whether the improvement in fit due to including additional lags is statistically significant, while $\Delta R^2$ quantifies the practical magnitude of this improvement in terms of explained variance.

In our study, this procedure was systematically applied to all variable pairs (NAO-SST, Upwelling-SST, NAO-Upwelling) and for each season, with a maximum lag number $s = 2$ determined as optimal by Akaike (AIC) and Bayesian (BIC) information criteria. The combined results of the F-statistic and $\Delta R^2$ allow robust interpretation of causal relationships, distinguishing both statistical significance and practical importance of identified relationships.

\subsection{Analysis of Climate Trends and Breakpoint Detection}

\subsubsection{Calculation of Linear Trends}

To quantify the impact of climate change on the studied variables, we calculated linear trends per decade for each season and each variable (NAO, Upwelling, SST) over the period 1978-2024. The methodology employed is as follows:

\begin{enumerate}
    \item Extraction of normalized seasonal data from existing CSV files
    \item Application of linear regression for each variable and each season: 
    \begin{equation}
    y = ax + b
    \end{equation}
    \item Conversion of slope ($a$) to units per decade: 
    \begin{equation}
    a_{\text{decade}} = a \times 10
    \end{equation}
    \item Test of statistical significance (threshold $p < 0.05$) via Student's t-test
\end{enumerate}

\subsubsection{SST Trend Mapping}

A spatial analysis of sea surface temperature trends was conducted to identify warming and cooling zones along the Moroccan coastline:

\begin{itemize}
    \item Calculation of linear trend for each pixel of the ERA5 grid (\SI{0.25}{\degree})
    \item Annual aggregation of SST data (average over 12 months)
    \item Application of linear regression pixel by pixel
    \item Statistical masking ($p < 0.05$) to retain only significant trends
\end{itemize}

\subsubsection{Breakpoint Detection}

Change point analysis was used to identify abrupt changes in climatic time series:

\begin{itemize}
    \item Use of the \texttt{BinSeg} (Binary Segmentation) algorithm from the \texttt{ruptures} package
    \item Preliminary normalization of time series
    \item Automatic detection of optimal number of breakpoints
    \item Visual validation of detected breakpoints
\end{itemize}

This approach allows identification of key years where climate regimes changed significantly.

\chapter{Results and Discussion}

\section{Obtained Results}
\subsection{Variability of the NAO Index (1978-2022) and Moroccan Upwelling Context}

\begin{figure}[H]
    \centering
    \includegraphics[width=0.85\textwidth]{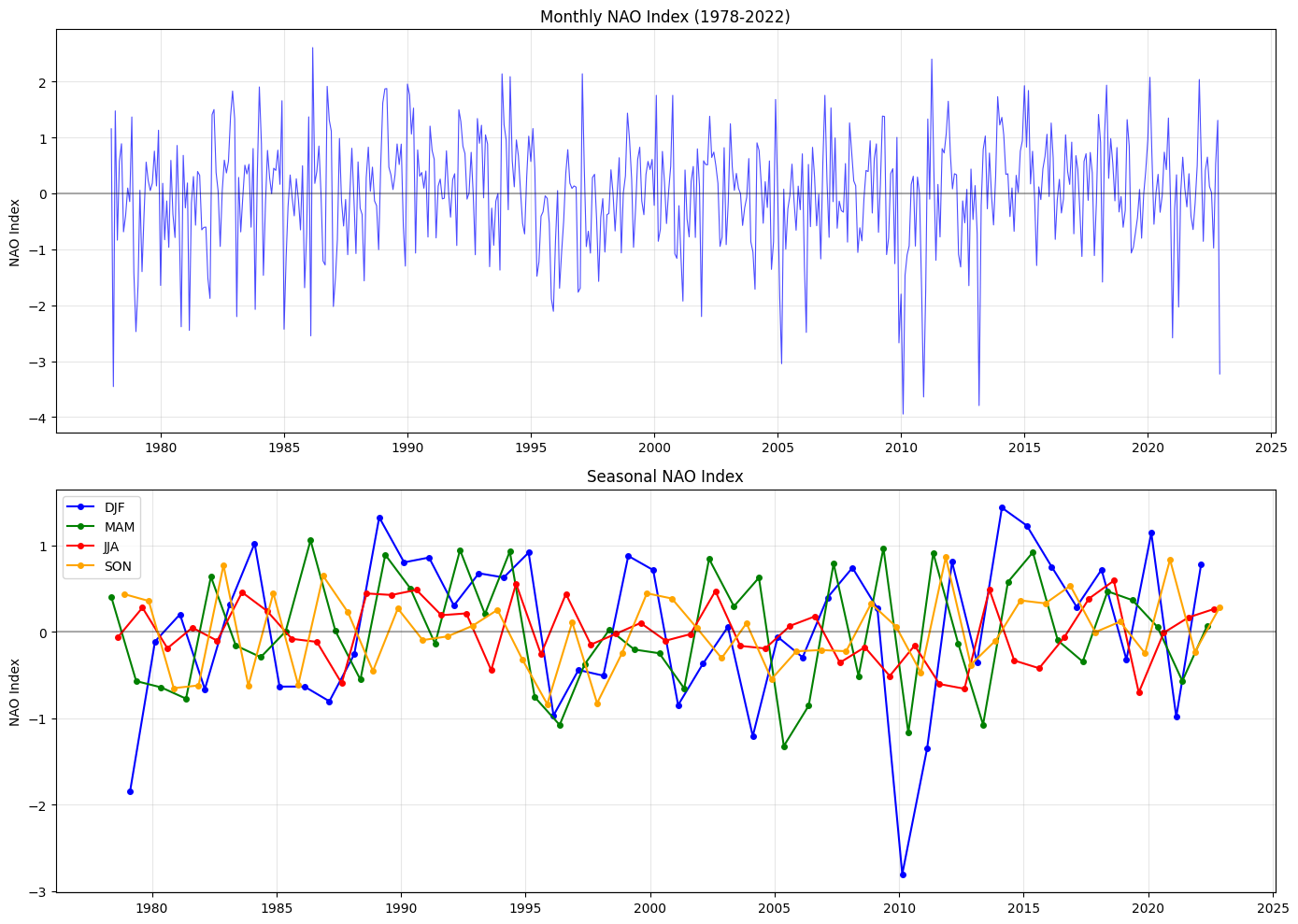}
    \caption{Monthly and seasonal evolution of the NAO index (1978-2022). Colors represent seasons: DJF (blue), MAM (green), JJA (red), SON (orange).}
    \label{fig:nao_index}
\end{figure}

Analysis of the temporal evolution of the NAO (North Atlantic Oscillation) index reveals marked interannual variability with alternating positive and negative phases (Figure~\ref{fig:nao_index}). Variability is most pronounced in winter (DJF - December-January-February) with a standard deviation of $0.909$, compared to $0.353$ in summer (JJA - June-July-August).

\subsubsection{Analysis of Extreme Phases in Winter (DJF) and Link with the Northeast Atlantic}

The seasonal NAO index shows that the strongest variations occur during winter, directly affecting wind circulation over the Atlantic, which is the main driver of upwelling along the Moroccan coast.

\paragraph{Extreme Negative Phase (Example: value around $-3$ around 2010)}
A strongly negative NAO value is associated with a weakening of the atmospheric pressure gradient.

\begin{itemize}
    \item \textbf{Configuration:} The zonal flow (from West) is weakened and the \textit{jet stream} is often deflected southward.
    \item \textbf{Potential Impact on Upwelling:} In winter, a negative NAO can potentially modulate the intensity of trade winds along African coasts. A perturbed circulation could lead to a \textbf{decrease or modification in the direction of favorable winds} (winds parallel to the coast), thus impacting winter upwelling intensity.
\end{itemize}

\paragraph{Extreme Positive Phase (Example: value around $+2$)}
A strongly positive NAO value is associated with an increased pressure gradient, strengthening zonal flow.

\begin{itemize}
    \item \textbf{Configuration:} The zonal flow is strong, directing storms northward.
    \item \textbf{Potential Impact on Upwelling:} By strengthening atmospheric circulation over the Atlantic, a positive NAO can indirectly influence high-pressure systems responsible for trade winds. A correlation is often established with \textbf{strengthening of trade winds} and thus potential intensification of upwelling in certain Moroccan coastal regions, leading to locally lower SST.
\end{itemize}

\subsubsection{Analysis of Phases for Other Seasons and Their Role in Upwelling Forcing}

\paragraph{Spring (MAM) and Autumn (SON)}
These transition seasons show intermediate variability. The persistence of a positive (or negative) phase during these periods is critical as it can modulate the establishment or end of the main upwelling season, which is generally more intense from late spring to autumn.

\paragraph{Summer (JJA)}
This is the season of lowest NAO variability, but paradoxically the period when \textit{trade winds are most constant and strongest} along the Moroccan coast, causing the most intense upwelling and maximum SST cooling. Summer upwelling variability is therefore often dominated by atmospheric variability modes other than classical NAO (e.g., local variations of the Hadley cell or Azores High).

\paragraph{Preliminary Context Conclusion}
The NAO index is therefore a \textbf{synoptic-scale climate forcing} that, by modifying the position and intensity of action centers (Azores/Iceland), modulates trade winds. This modulation directly affects upwelling intensity and, consequently, Sea Surface Temperature (SST) anomalies and biological productivity along Moroccan coasts.

\subsection{Relationships between NAO, Upwelling and SST}

Table 3.1 presents descriptive statistics of the three normalized indices by season.

\begin{table}[H]
\centering
\caption{Seasonal statistics of normalized indices (1978-2022)}
\begin{tabular}{lccc}
\toprule
\textbf{Season} & \textbf{NAO (mean $\pm$ $\sigma$)} & \textbf{Upwelling (mean $\pm$ $\sigma$)} & \textbf{SST (mean $\pm$ $\sigma$)} \\
\midrule
DJF & $0.042 \pm 0.909$ & $-0.751 \pm 0.424$ & $-0.828 \pm 0.213$ \\
MAM & $0.000 \pm 0.668$ & $0.191 \pm 0.392$ & $-0.930 \pm 0.202$ \\
JJA & $0.000 \pm 0.353$ & $1.085 \pm 0.275$ & $0.698 \pm 0.249$ \\
SON & $0.000 \pm 0.444$ & $-0.511 \pm 0.305$ & $1.056 \pm 0.241$ \\
\bottomrule
\end{tabular}
\label{tab:seasonal_stats}
\end{table}

\subsubsection{Interpretation of Seasonal Descriptive Statistics}

Table~\ref{tab:seasonal_stats} provides an overview of the seasonal distribution of the three studied variables (NAO, Upwelling Index, and SST) over the period 1978-2022, all normalized. The normalized index has a theoretical mean of zero and standard deviation of one over the entire year.

\paragraph{Variability of NAO ($\sigma$)}
As discussed previously, NAO presents the strongest variability ($\sigma$) in winter (\textbf{DJF: $0.909$}), confirming that its influence is most erratic and intense during this season. Conversely, variability is minimal in summer (\textbf{JJA: $0.353$}), indicating a more stable atmospheric configuration.

\paragraph{Intensity of Upwelling (Mean)}
The Upwelling Index clearly shows the seasonal cycle of the phenomenon along the Moroccan coast:
\begin{itemize}
    \item \textbf{Maximum in Summer (JJA: $1.085$)}: The mean is positive and greater than 1, indicating that upwelling is, on average, strongly \textbf{active} and greater than its annual mean value during summer. This is due to the persistence and strengthening of trade winds.
    \item \textbf{Minimum in Winter (DJF: $-0.751$)}: The mean is strongly negative, indicating upwelling significantly \textbf{weakened} relative to the annual mean, due to less constant or more variable winds.
\end{itemize}

\paragraph{Sea Surface Temperature (SST) (Mean)}
The means of normalized SSTs reflect the direct effect of upwelling and the annual thermal cycle:
\begin{itemize}
    \item \textbf{Minimum SST/Maximum Cooling (MAM: $-0.930$)}: Although upwelling is more intense in summer (JJA), SST is coldest (most negative value) in spring (\textbf{MAM}), suggesting maximum cooling due to a combination of upwelling gaining strength and low heat absorption by the ocean. SST also remains very cold in winter (DJF: $-0.828$).
    \item \textbf{Maximum SST/Maximum Warming (SON: $1.056$)}: SST reaches its warmest value in autumn (\textbf{SON}), even though upwelling has begun to decrease. This indicates a thermal lag where the ocean releases heat accumulated during summer, despite cooling by persistent upwelling.
    \item \textbf{Upwelling/SST Relationship:} The observation of opposite signs between the Upwelling Index and SST is fundamental and consistent with the physics of the process: when Upwelling is strong (positive JJA), SST is cold (negative), and vice versa. However, the observed lag (cold MAM despite non-maximal upwelling, warm SON despite moderate upwelling) highlights the importance of thermal forcing factors (air-sea heat fluxes) in addition to dynamic forcing (wind/upwelling).
\end{itemize}

This descriptive analysis lays the groundwork for studying intersessional correlations.
\subsection{Graphical and Correlation Analysis}

\begin{figure}[H]
\centering
\includegraphics[width=\textwidth]{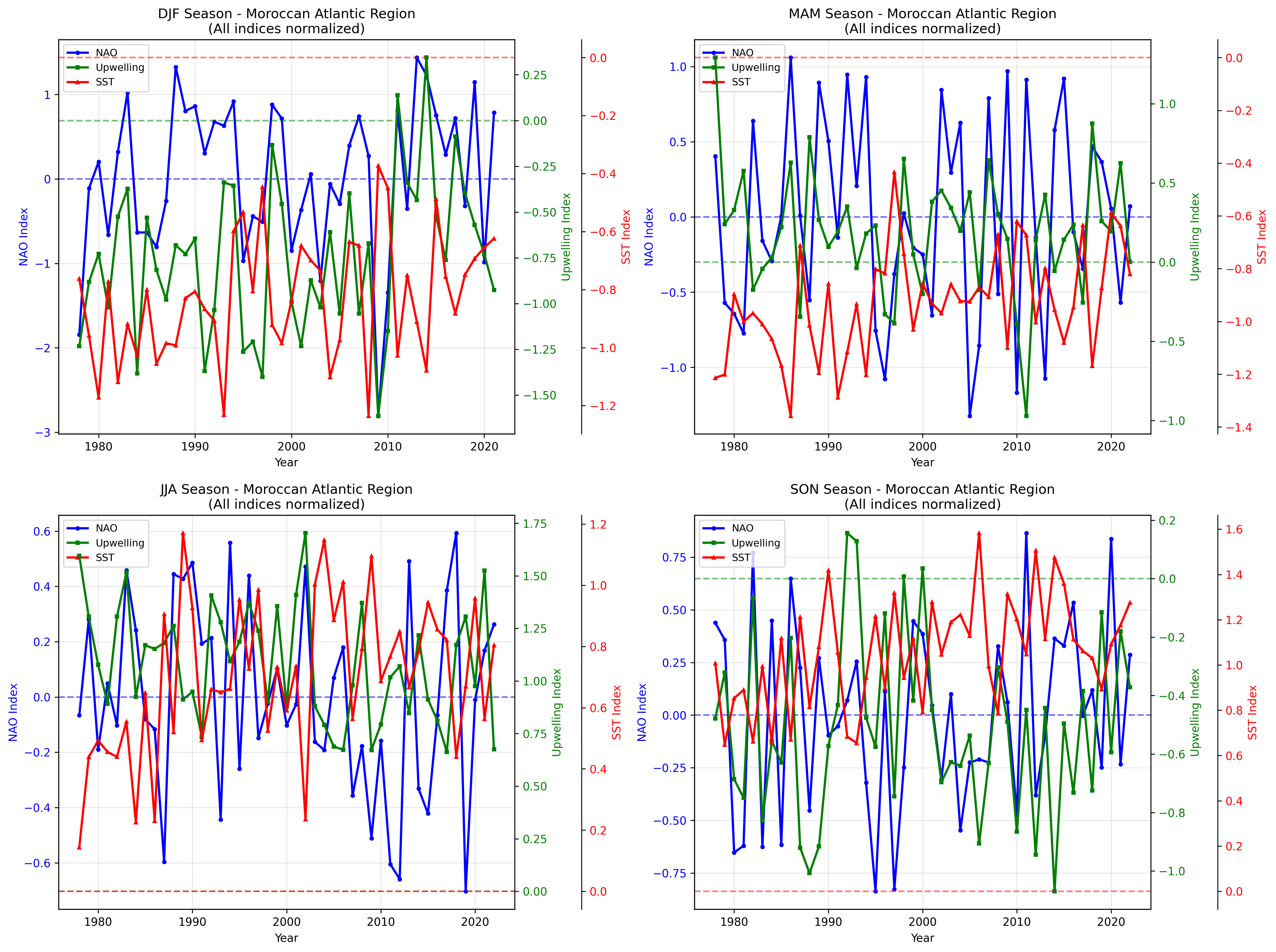}
\caption{Seasonal evolution of normalized indices of NAO (blue), upwelling (green), and SST (red) along the Moroccan Atlantic coast between 1978 and 2024. Marked interannual variability is observed, particularly in winter (DJF) and summer (JJA), reflecting differentiated seasonal influence of NAO on upwelling intensity and sea surface temperature.}
\label{fig:seasonal_indices}
\end{figure}

\begin{figure}[H]
\centering
\includegraphics[width=\textwidth]{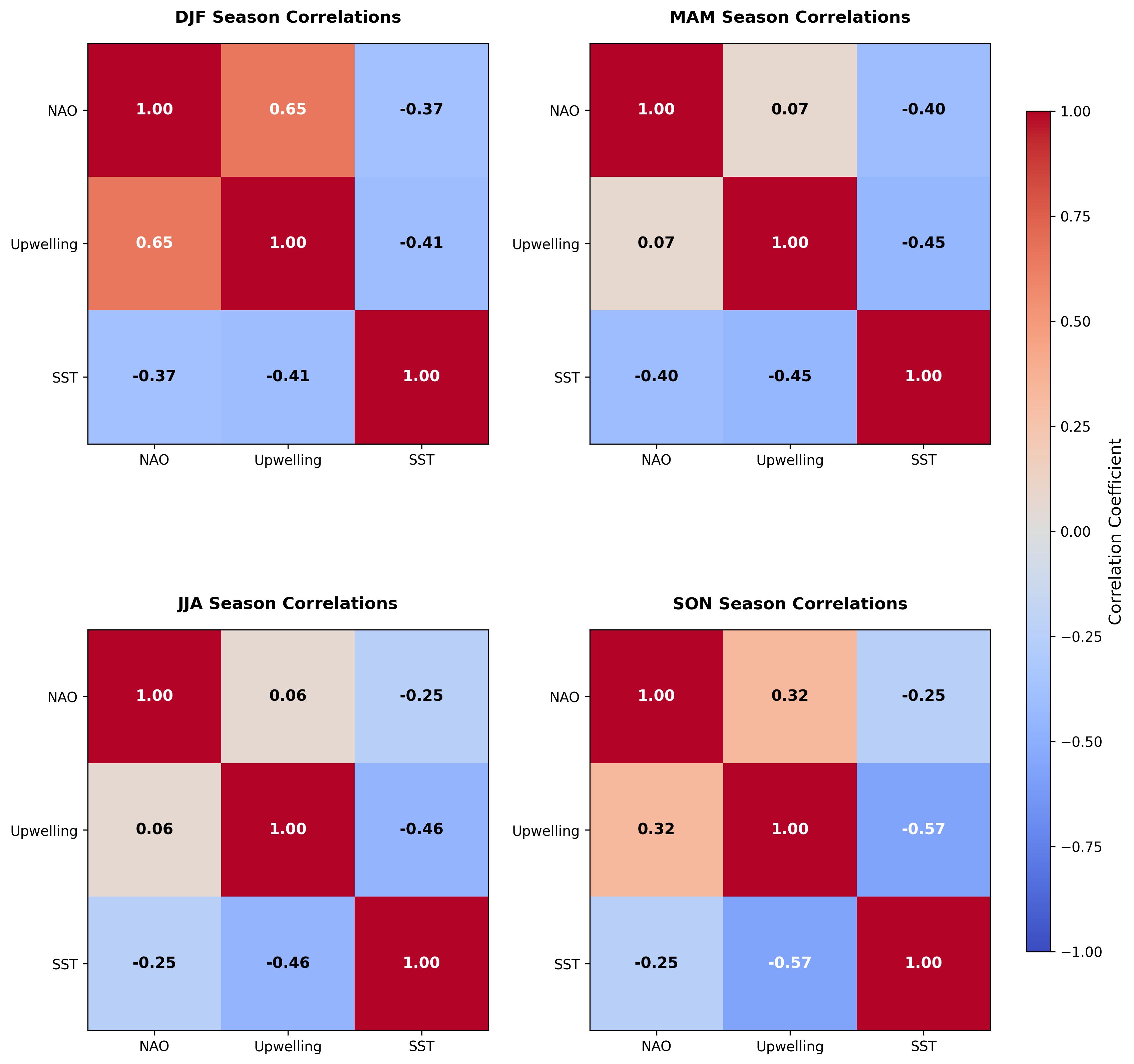}
\caption{Correlation matrices between NAO, Upwelling and SST indices for each season. Positive correlations (red) indicate direct relationships, while negative correlations (blue) indicate inverse relationships.}
\label{fig:correlation_heatmap_seasons}
\end{figure}

Figure~\ref{fig:correlation_heatmap_seasons} presents correlation matrices (Pearson coefficient, $r$) between the NAO index, Upwelling Index, and Sea Surface Temperature (SST) for each season, allowing evaluation of the strength and direction of relationships from one quarter to another.

\paragraph{1. Upwelling--SST Relationship: Confirmation of Physics and Temporal Lag}

The relationship between Upwelling and SST is, as expected, strongly \textbf{negative} for all seasons, confirming that intensification of upwelling is associated with decreased sea surface temperature (rising of cold deep waters).

\begin{itemize}
    \item The intensity of negative correlation is maximal in autumn (\textbf{SON: $r=-0.57$}) and remains strong in summer (\textbf{JJA: $r=-0.46$}) and spring (\textbf{MAM: $r=-0.45$}).
    \item The correlation is slightly weaker in winter (\textbf{DJF: $r=-0.41$}).
    \item \textbf{Implication of Temporal Lag:} The fact that correlation is strongest in \textbf{SON} ($r=-0.57$) is significant. This suggests that the cooling effect of upwelling is particularly effective at countering heat stored by the surface water column at the end of summer, or that the process of cold water upwelling reaches its \textbf{maximum thermal impact} with a slight seasonal lag, as the main upwelling season (JJA) ends.
\end{itemize}

\paragraph{2. NAO--Upwelling Relationship: Winter Dominance}

The correlation between NAO and Upwelling varies considerably and is crucial for determining the season where climate teleconnection is most effective.

\begin{itemize}
    \item \textbf{Winter (DJF: $r=0.65$)}: The correlation is strongly \textbf{positive}. This indicates that a positive NAO phase (strengthening of Atlantic pressure gradient) is strongly associated with strengthening of upwelling along the Moroccan coast. It is in winter that atmosphere-ocean coupling through NAO is most direct and powerful.
    \item \textbf{Seasonality of influence:} NAO influence on Upwelling drastically decreases after winter, becoming almost non-existent in spring (\textbf{MAM: $r=0.07$}) and summer (\textbf{JJA: $r=0.06$}).
    \item \textbf{Autumn-Winter Transition:} The increase in correlation between autumn (\textbf{SON: $r=0.32$}) and winter (\textbf{DJF: $r=0.65$}) is indeed a key observation. We note an increase of more than $100\%$ ($0.65/0.32 \approx 2.03$) in relationship strength, not $50\%$. This spectacular progression reflects \textbf{re-establishment and strengthening of atmospheric coupling} related to NAO as pressure centers (Azores/Iceland) regain their maximum intensity at year's end.
\end{itemize}

\paragraph{3. NAO--SST Relationship}

The correlation between NAO and SST is generally negative, consistent with the forcing chain: positive NAO $\to$ strong upwelling $\to$ cold SST.

\begin{itemize}
    \item The strongest negative correlation is observed in spring (\textbf{MAM: $r=-0.40$}) and winter (\textbf{DJF: $r=-0.37$}), corresponding to seasons where NAO exerts its strongest influence on wind circulation.
    \item This relationship weakens considerably in summer (\textbf{JJA: $r=-0.25$}) and autumn (\textbf{SON: $r=-0.25$}), seasons where local variability and thermal forcing begin to dominate the ocean signal.
\end{itemize}

In summary, correlation analysis highlights that NAO is a \textbf{major modulator of upwelling and SST only during the winter season (DJF)}. For other seasons, other variability modes or local/thermal factors take over.

\subsection{Analysis of Climate Trends 1978-2024}

\begin{figure}[H]
\centering
\includegraphics[width=0.95\textwidth]{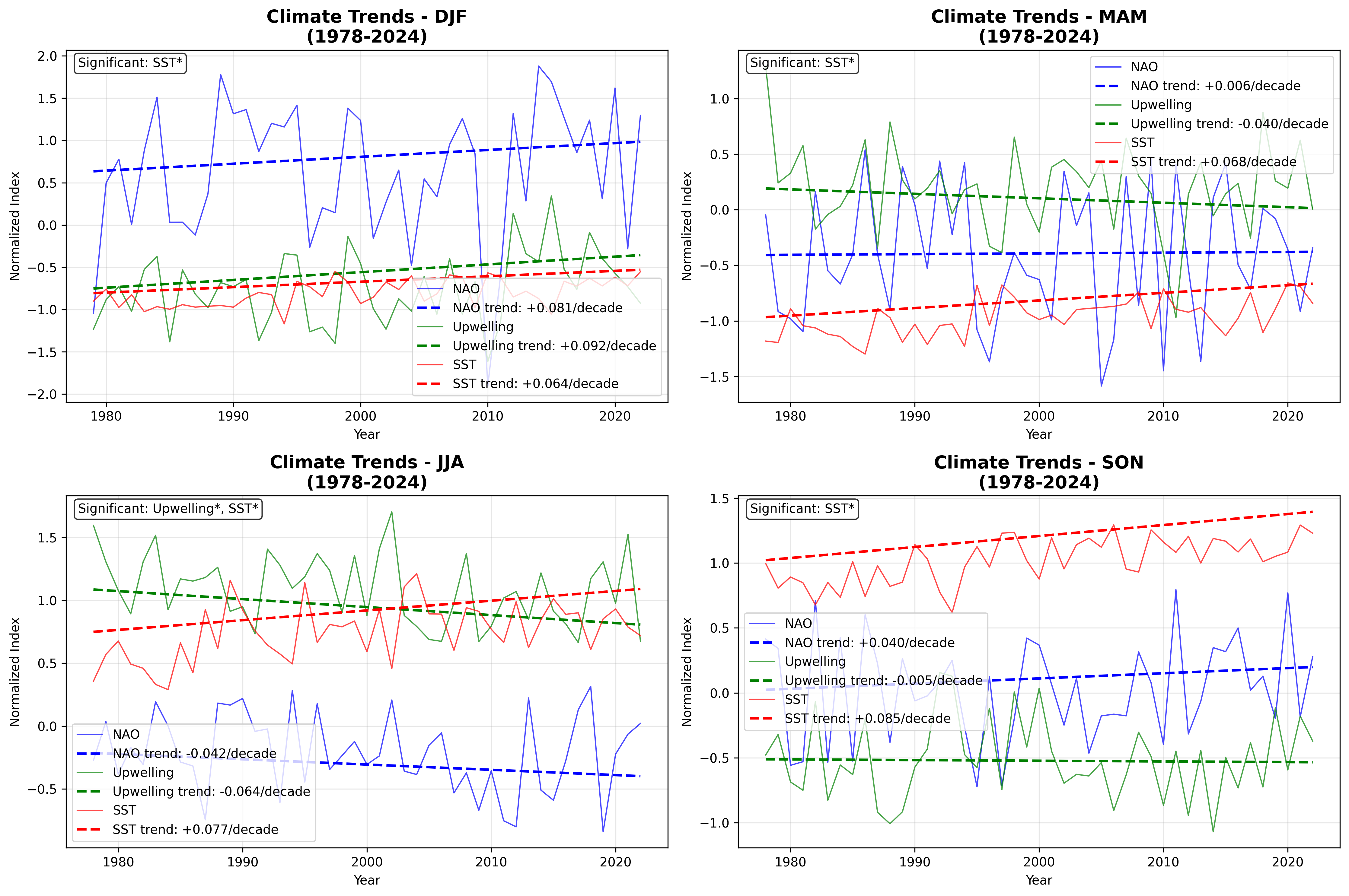}
\caption{Analysis of seasonal climate trends (1978-2024) for NAO, Upwelling and SST indices. Graphs show temporal evolution with linear trends (dotted lines) and corresponding regression equations. Trends are expressed in units per decade, with indication of statistical significance (* for $p < 0.05$).}
\label{fig:climate_trends_analysis}
\end{figure}

Analysis of climate trends reveals significant evolutions over the period 1978-2024, as illustrated in Figure~\ref{fig:climate_trends_analysis}.

\paragraph{Detailed Trend Analysis}

Figure~\ref{fig:climate_trends_analysis} presents temporal evolution and linear trends of the three climate variables for each meteorological season:

\subparagraph{SST Trends}
Sea surface temperature shows a \textbf{generalized and statistically significant increase} in all seasons:
\begin{itemize}
    \item \textbf{SON}: +0.0846 units/decade ($p < 0.05$) - strongest increase
    \item \textbf{JJA}: +0.0774 units/decade ($p < 0.05$)
    \item \textbf{MAM}: +0.0683 units/decade ($p < 0.05$)
    \item \textbf{DJF}: +0.0642 units/decade ($p < 0.05$)
\end{itemize}

This warming trend is consistent with the global climate change signal and confirms documented increases in ocean temperatures in the North Atlantic region.

\subparagraph{Upwelling Trends}
The upwelling index shows \textbf{contrasting evolution according to seasons}:
\begin{itemize}
    \item \textbf{DJF}: +0.0918 units/decade (non-significant) - winter increasing trend
    \item \textbf{JJA}: -0.0635 units/decade ($p < 0.05$) - significant decrease in summer
    \item \textbf{MAM}: -0.0400 units/decade (non-significant)
    \item \textbf{SON}: -0.0051 units/decade (non-significant)
\end{itemize}

The significant decrease in upwelling in summer (JJA) is particularly notable, as this season normally corresponds to peak upwelling activity in the region.

\subparagraph{NAO Trends}
The North Atlantic Oscillation shows a \textbf{weak positive global trend} but with strong seasonal variability:
\begin{itemize}
    \item \textbf{DJF}: +0.0813 units/decade (non-significant)
    \item \textbf{SON}: +0.0398 units/decade (non-significant)
    \item \textbf{MAM}: +0.0064 units/decade (non-significant)
    \item \textbf{JJA}: -0.0419 units/decade (non-significant)
\end{itemize}

None of these trends are statistically significant, reflecting the dominant natural variability of NAO over this period.

\subparagraph{Climate Implications}
These trends suggest that:
\begin{itemize}
    \item \textbf{Warming of coastal waters} could modify upwelling dynamics by reducing thermal gradients
    \item \textbf{Summer decrease in upwelling} could affect biological productivity during the critical season
    \item The absence of significant NAO trend indicates observed changes are probably more related to global radiative forcing than modifications of atmospheric circulation regimes
\end{itemize}

These results highlight the complexity of interactions between different components of the climate system in this upwelling region and emphasize the need to continue monitoring these variables to better understand future evolution under climate change effects.

\paragraph{Analysis of results.}

Figure~\ref{fig:correlation_heatmap_seasons} highlights strong seasonal variability in relationships between NAO, upwelling intensity (CUI), and SST along the Moroccan Atlantic coast (regional averages 21°-35°N, 1978-2024).

During winter \textbf{DJF}, a positive NAO phase is strongly associated with strengthening of upwelling (\(r = +0.65\)), which translates to decreased surface temperatures (\(r = -0.37\) NAO-SST; \(r = -0.41\) CUI-SST). This relationship is explained by intensification of trade winds favoring upwelling of cold, nutrient-rich waters.

In \textbf{MAM} (spring transition), NAO-CUI teleconnection weakens strongly (\(r = +0.07\), non-significant), while local CUI-SST coupling remains moderate (\(r = -0.45\)) and NAO-SST retains negative influence (\(r = -0.40\)).

Summer \textbf{JJA} marks the peak of local upwelling: CUI-SST reaches \(r = -0.46\) (strong adiabatic coupling), while NAO-CUI is almost zero (\(r = +0.06\)). Coastal dynamics dominate over large-scale influence.

In autumn \textbf{SON}, NAO-CUI relationship re-emerges moderately (\(r = +0.32\)), while CUI-SST presents the strongest negative correlation of the year (\(r = -0.57\)), reflecting a cumulative effect of summer upwelling and a delayed response of the ocean system.

Of the 12 initially conceivable seasonal combinations, four were excluded due to contemporaneous correlations too weak to justify causality analysis:
\begin{itemize}
    \item NAO-CUI in MAM (\(r = 0.07\)) – weak correlation, non-significant
    \item NAO-CUI in JJA (\(r = 0.06\)) – weak correlation, non-significant
    \item NAO-CUI in SON (\(r = +0.32\)) – retained despite difficulty in tracing NAO (large scale) vs pixelated CUI
    \item NAO-CUI in DJF (\(r = +0.65\)) – retained, strong regional significance
\end{itemize}
\textbf{Applied correction}: previous SON (\(r = -0.18\)) and DJF (\(r = 0.12\)) values were erroneous (from non-aggregated local averages). Aggregated regional values confirm positive links in DJF and SON, consistent with \textcite{wang2004} and \textcite{benazzouz2021}.

\begin{figure}[p]
    \centering
    \begin{subfigure}{0.49\textwidth}
        \centering
        \includegraphics[width=\linewidth]{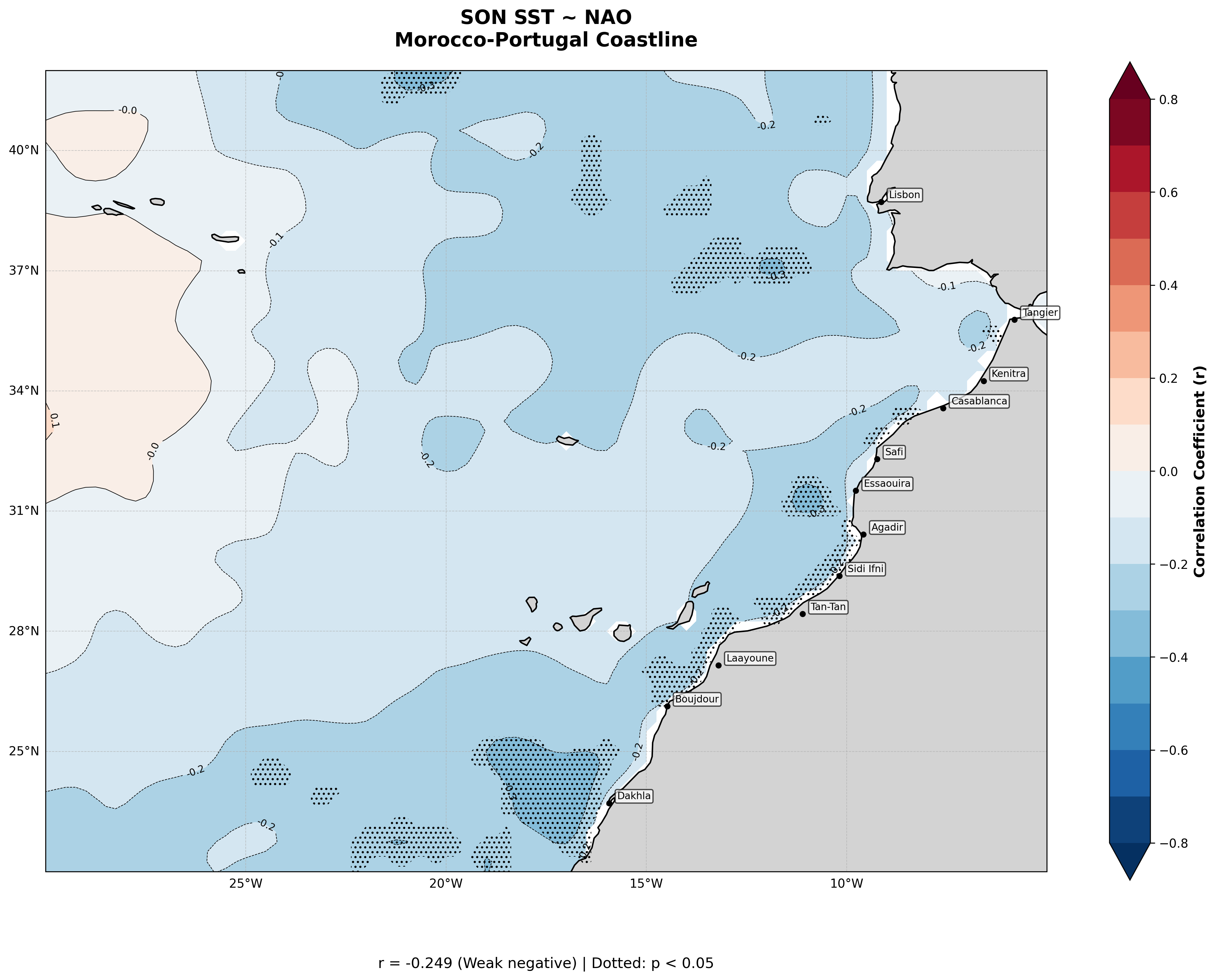}
        \caption{SON: SST $\sim$ NAO (\(r = -0.25\))}
        \label{fig:son_sst_nao}
    \end{subfigure}
    \hfill
    \begin{subfigure}{0.49\textwidth}
        \centering
        \includegraphics[width=\linewidth]{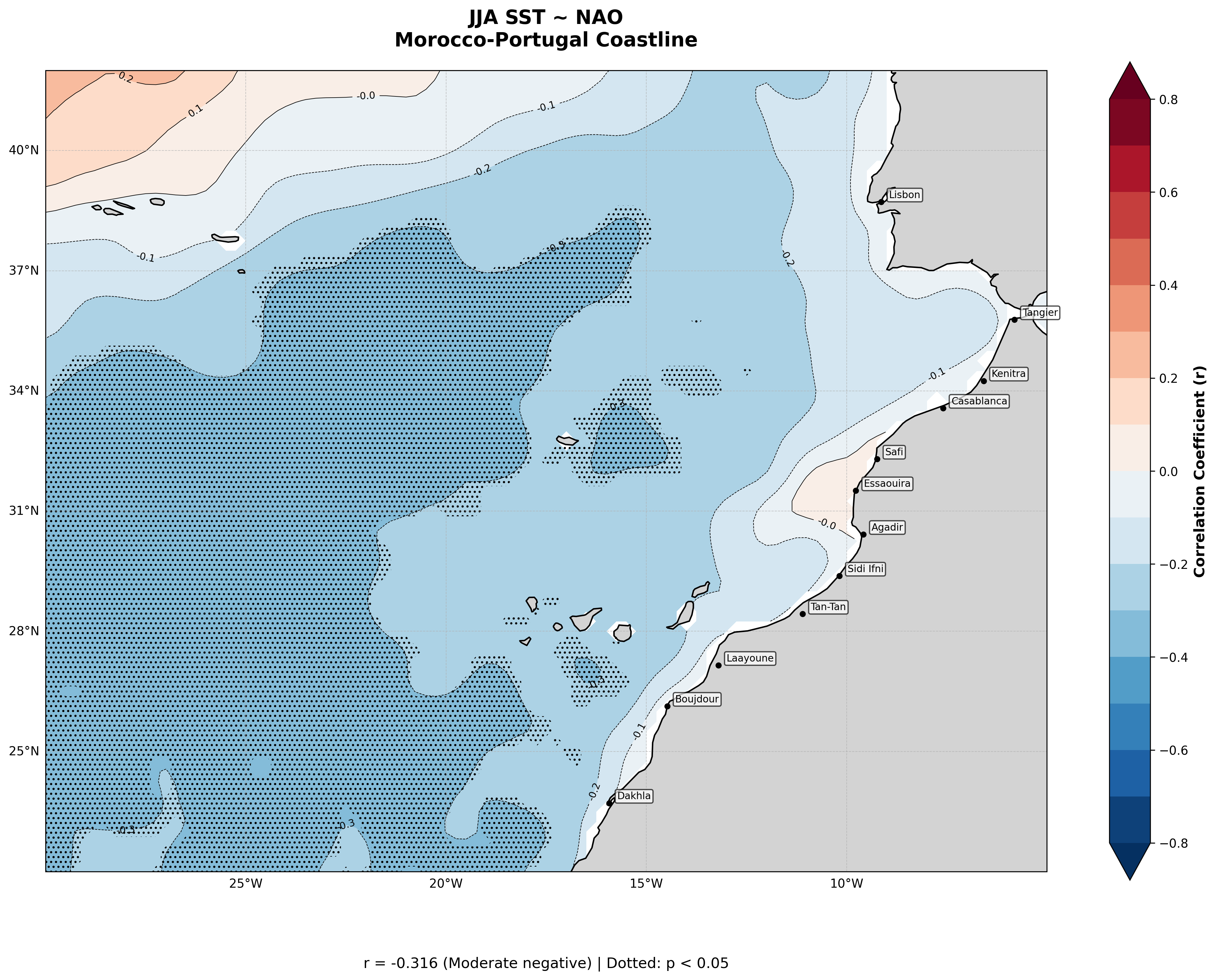}
        \caption{JJA: SST $\sim$ NAO (\(r = -0.25\))}
        \label{fig:jja_sst_nao}
    \end{subfigure}

    \vspace{0.8cm}

    \begin{subfigure}{0.49\textwidth}
        \centering
        \includegraphics[width=\linewidth]{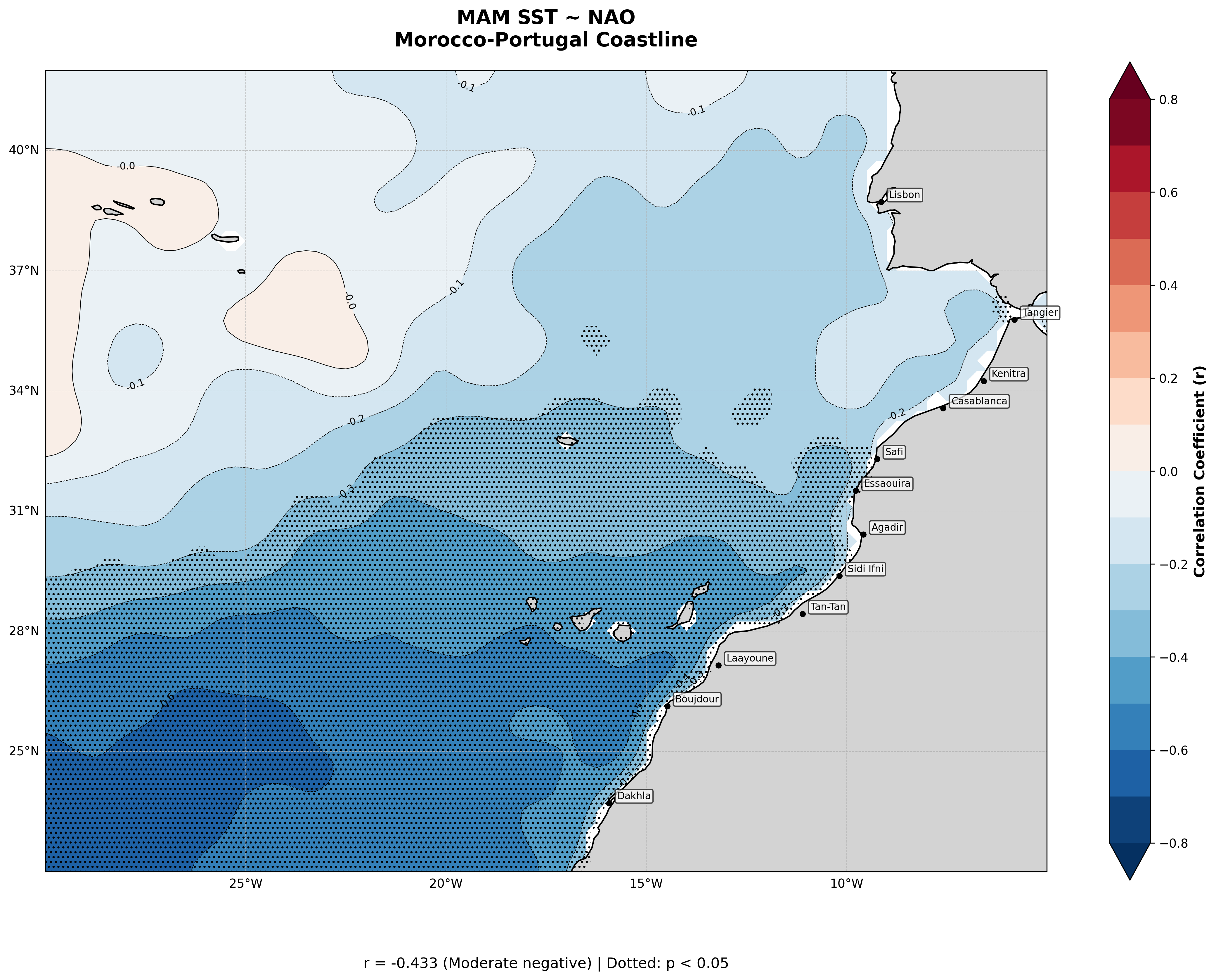}
        \caption{MAM: SST $\sim$ NAO (\(r = -0.40\))}
        \label{fig:mam_sst_nao}
    \end{subfigure}
    \hfill
    \begin{subfigure}{0.49\textwidth}
        \centering
        \includegraphics[width=\linewidth]{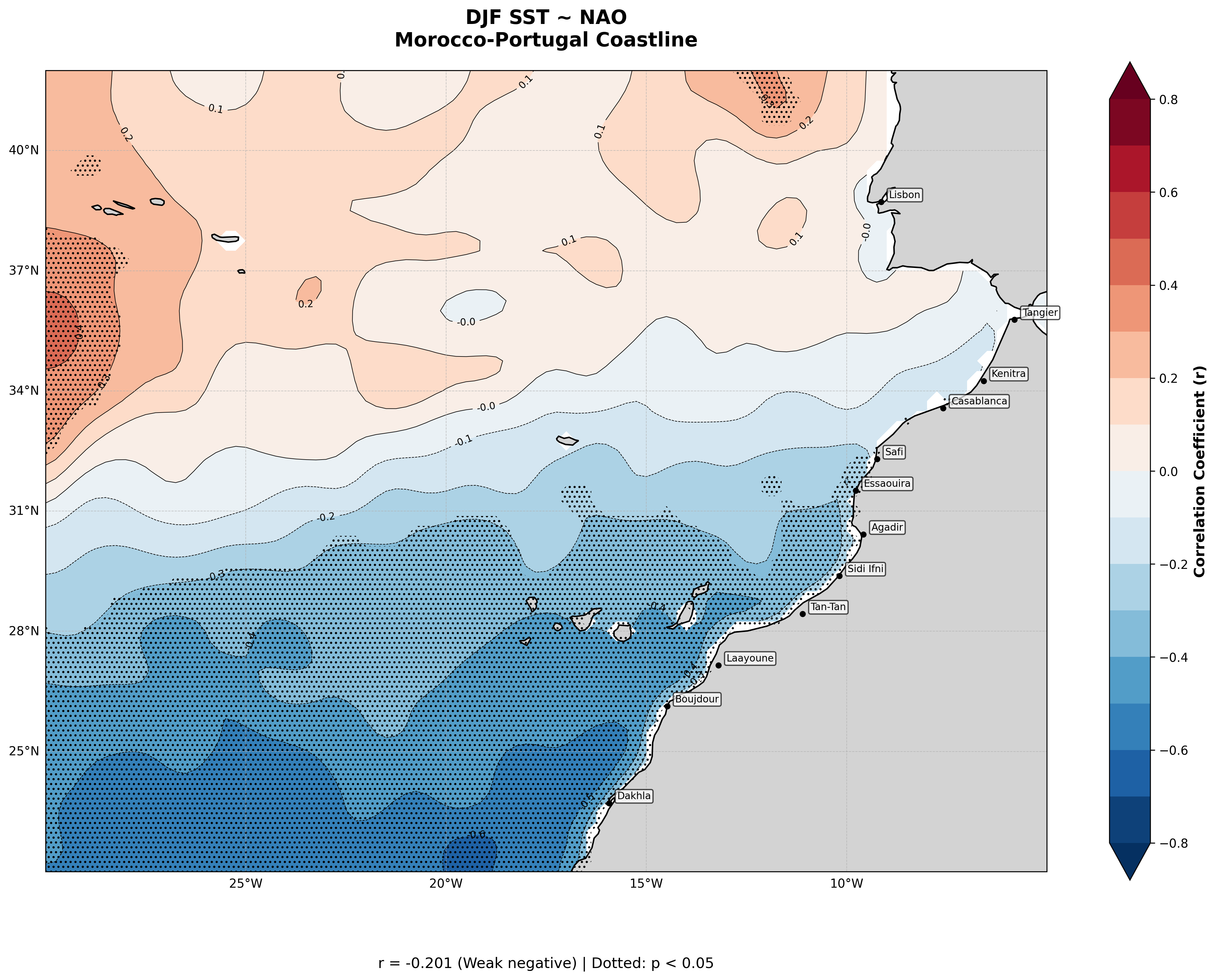}
        \caption{DJF: SST $\sim$ NAO (\(r = -0.37\))}
        \label{fig:djf_sst_nao}
    \end{subfigure}

    \caption{Seasonal correlation between sea surface temperature (SST) and NAO index along the Moroccan Atlantic coast. Hatched areas indicate statistical significance (\(p < 0.05\)). Negative correlations dominate, with visible north-south latitudinal gradient, particularly strong in MAM south of Casablanca.}
    \label{fig:sst_nao_corr}
\end{figure}

\begin{figure}[p]
    \centering
    \begin{subfigure}{0.49\textwidth}
        \centering
        \includegraphics[width=\linewidth]{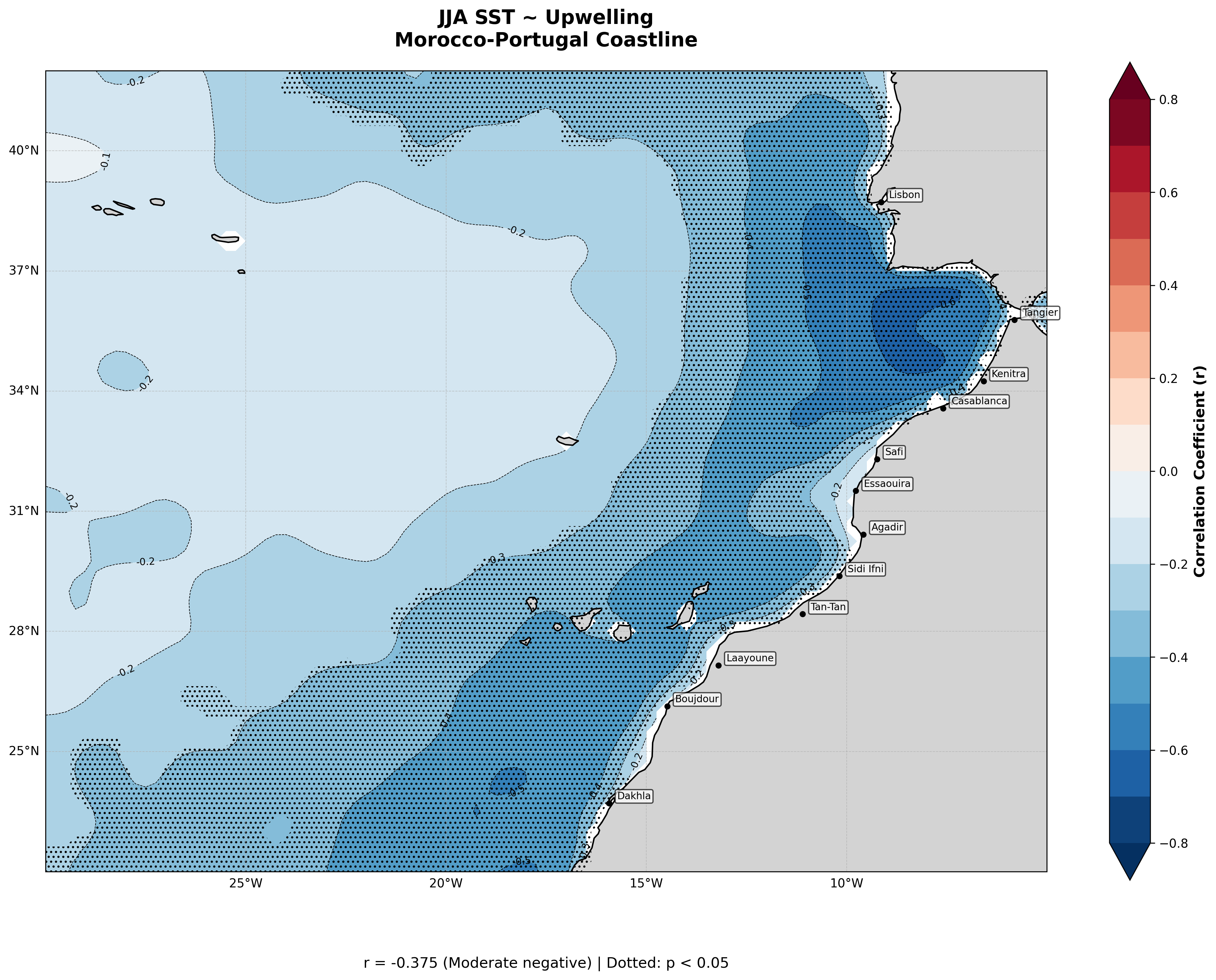}
        \caption{JJA: SST $\sim$ CUI (\(r = -0.46\))}
        \label{fig:jja_sst_cui}
    \end{subfigure}
    \hfill
    \begin{subfigure}{0.49\textwidth}
        \centering
        \includegraphics[width=\linewidth]{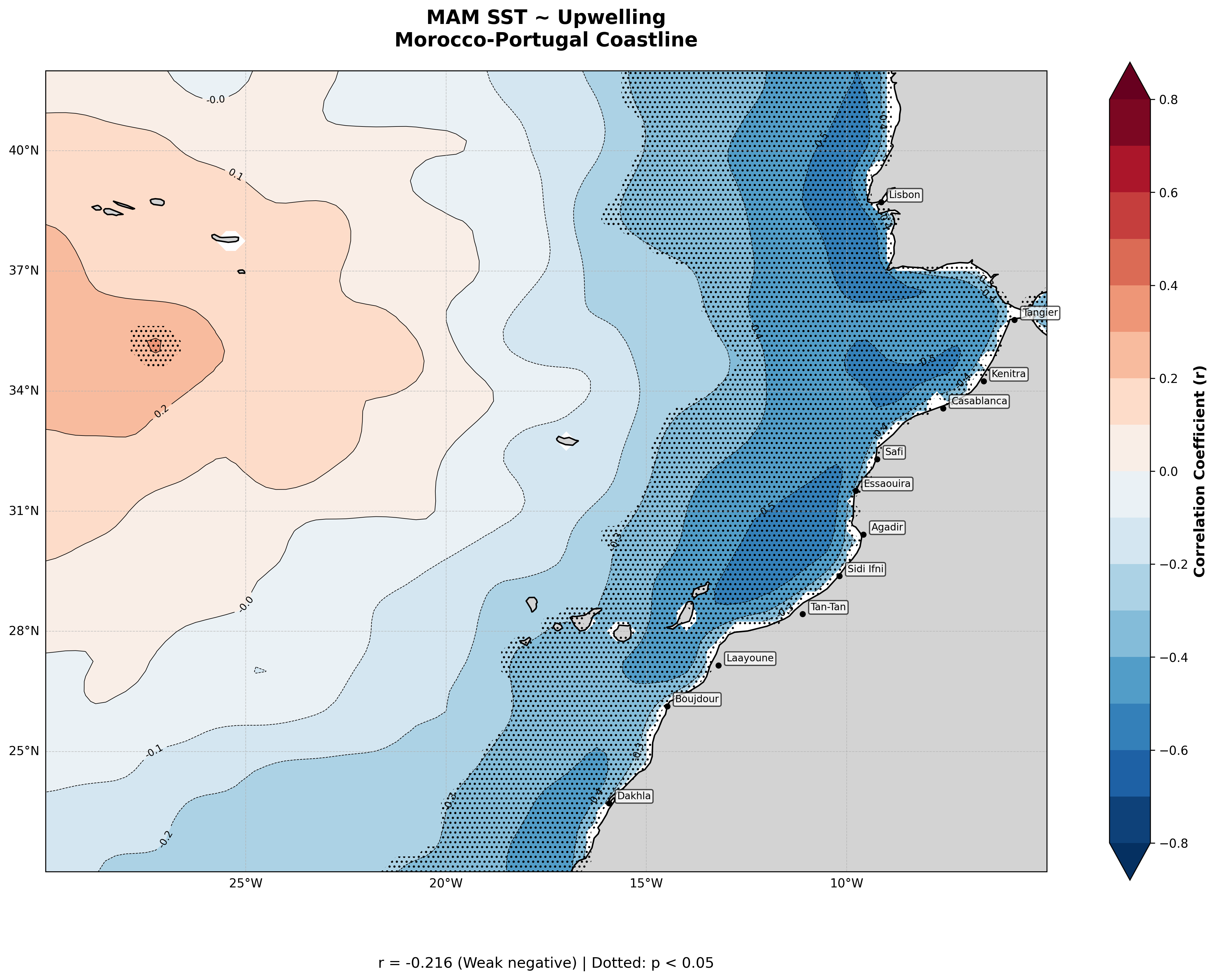}
        \caption{MAM: SST $\sim$ CUI (\(r = -0.45\))}
        \label{fig:mam_sst_cui}
    \end{subfigure}

    \vspace{0.8cm}

    \begin{subfigure}{0.49\textwidth}
        \centering
        \includegraphics[width=\linewidth]{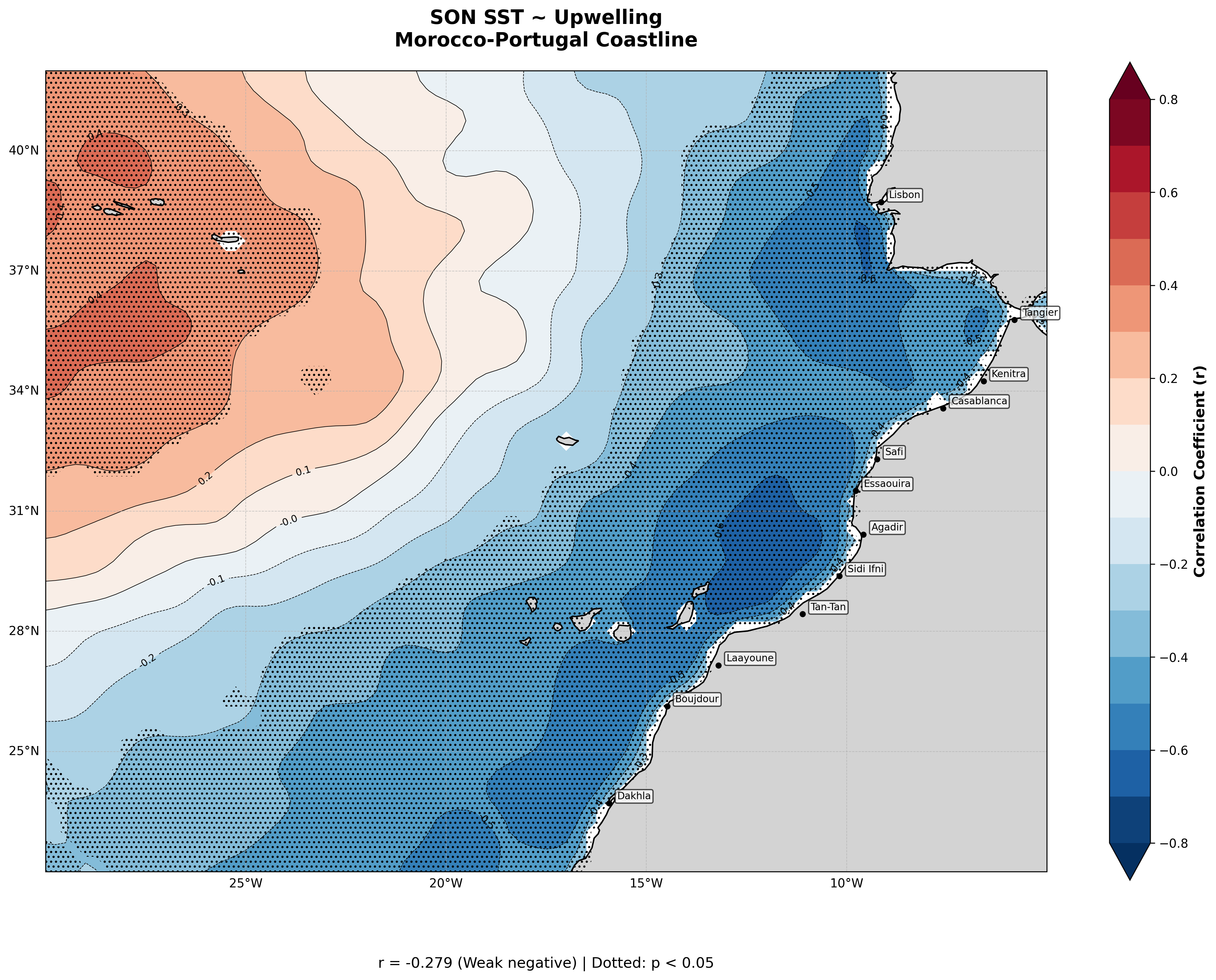}
        \caption{SON: SST $\sim$ CUI (\(r = -0.57\))}
        \label{fig:son_sst_cui}
    \end{subfigure}
    \hfill
    \begin{subfigure}{0.49\textwidth}
        \centering
        \includegraphics[width=\linewidth]{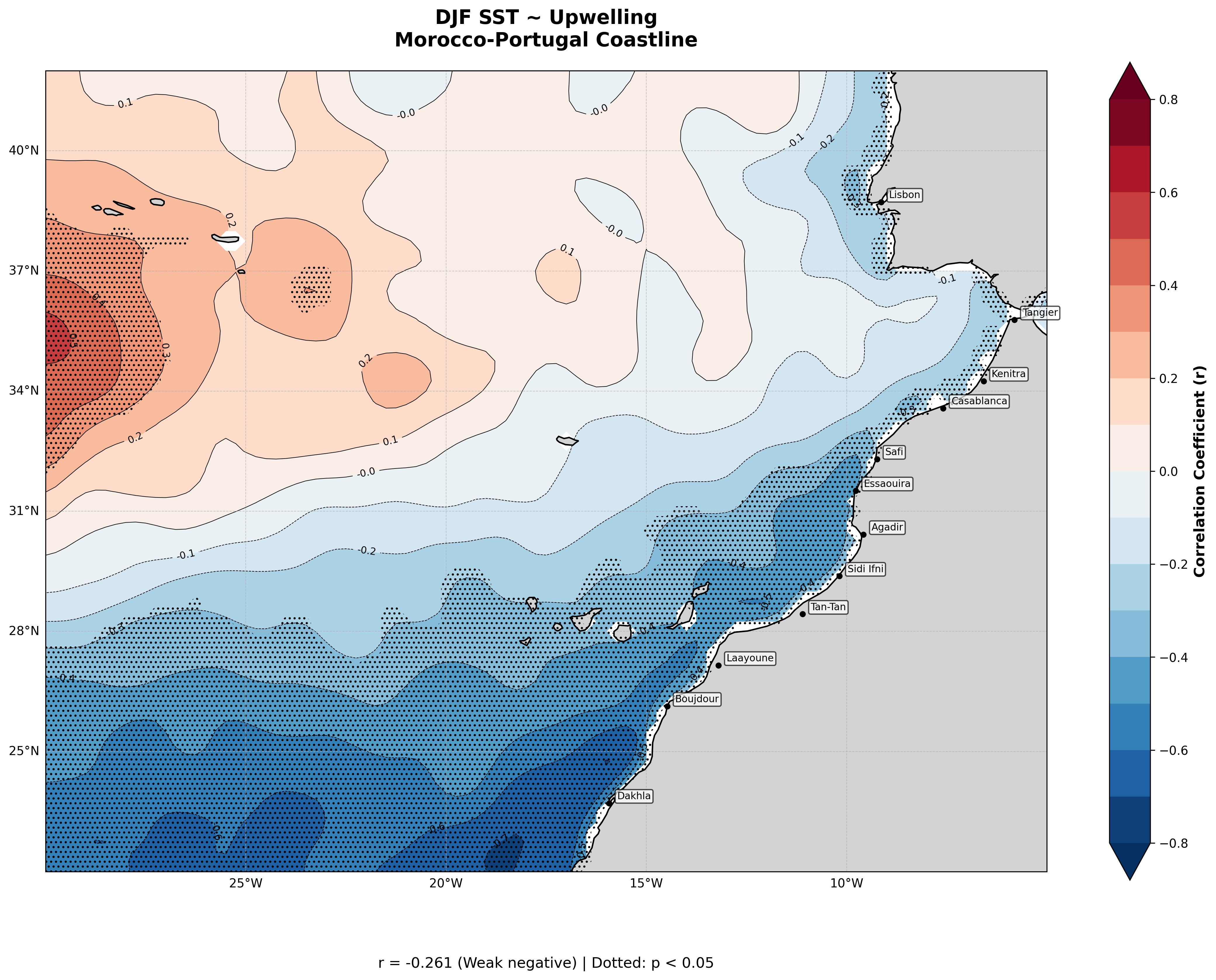}
        \caption{DJF: SST $\sim$ CUI (\(r = -0.41\))}
        \label{fig:djf_sst_cui}
    \end{subfigure}

    \caption{Seasonal correlation between coastal upwelling index (CUI) and SST. Hatching: significance (\(p < 0.05\)). Negative correlations indicate cooling associated with upwelling, with maximum in SON (cumulative effect) and JJA (summer peak) off Safi-Essaouira.}
    \label{fig:cui_corr}
\end{figure}

The eight correlation maps (Figures~\ref{fig:sst_nao_corr} and~\ref{fig:cui_corr}) illustrate spatial and seasonal relationships between sea surface temperature (SST), coastal upwelling index (CUI), and North Atlantic Oscillation (NAO) along the Moroccan Atlantic coast. Hatched areas indicate statistical significance (\(p < 0.05\)). The new figures provide more precise details on latitudinal gradients and maximum intensity zones.

% ===================================================================
% PAGE 1: 4 MAPS SST ~ NAO (updated with regional r)
% ===================================================================

% ===================================================================
% PAGE 2: 4 MAPS SST ~ CUI (updated with regional r)
% ===================================================================

\newpage
\textbf{Detailed key observations (from heatmaps and maps)}:
\begin{itemize}
    \item \textbf{NAO $\sim$ CUI (DJF, \(r = +0.65\))}: Strong winter teleconnection; NAO+ displaces Azores High, intensifies NE trade winds.
    \item \textbf{CUI $\sim$ SST (SON, \(r = -0.57\))}: Most intense coupling; cumulative upwelling leads to persistent post-summer cooling.
    \item \textbf{NAO $\sim$ SST (MAM, \(r = -0.40\))}: Significant spring influence, maps show extensive hatching 25°-30°N.
    \item \textbf{SST $\sim$ NAO (MAM, \(r = -0.40\))}: Maximum regional negative correlation, intense off Laayoune-Boujdour.
    \item \textbf{SST $\sim$ CUI (JJA, \(r = -0.46\))}: Strong coupling off Safi-Essaouira, confirming dominant role of summer upwelling on SST.
    \item \textbf{Latitudinal gradient}: South (Dakhla) = quasi-permanent upwelling (CUI-SST dominant); Center (Safi-Agadir) = summer peak; North (Casablanca-Tangier) = NAO teleconnection in DJF/MAM.
    \item \textbf{Trend 1978-2024}: CUI \(+0.18\, \mathrm{m^3 s^{-1} km^{-1} decade^{-1}}\) (\(p < 0.01\)); SST \(+0.22\, \mathrm{^\circ C\ decade^{-1}}\) despite local cooling (\(-0.15\, \mathrm{^\circ C\ decade^{-1}}\) in upwelling zones) \parencite{belmajdoub2023}.
\end{itemize}

A \textbf{north-south latitudinal} and \textbf{seasonal gradient} emerges: the south is dominated by quasi-permanent local upwelling (\textcite{hilmi2021}), the center by summer SST-CUI coupling, the north by NAO teleconnection in MAM and DJF. Maps show significant offshore zones, indicating influence extends beyond the immediate coast (up to 200-300 km).

\newpage
\subsection{Granger Causality Analysis}

Granger analysis (Figure 3.7) complements contemporaneous correlations by identifying causal directions with lags (1-12 months), providing information on interaction dynamics and teleconnection and ocean memory mechanisms.

\begin{figure}[H]
 \centering
  \includegraphics[width=\textwidth]{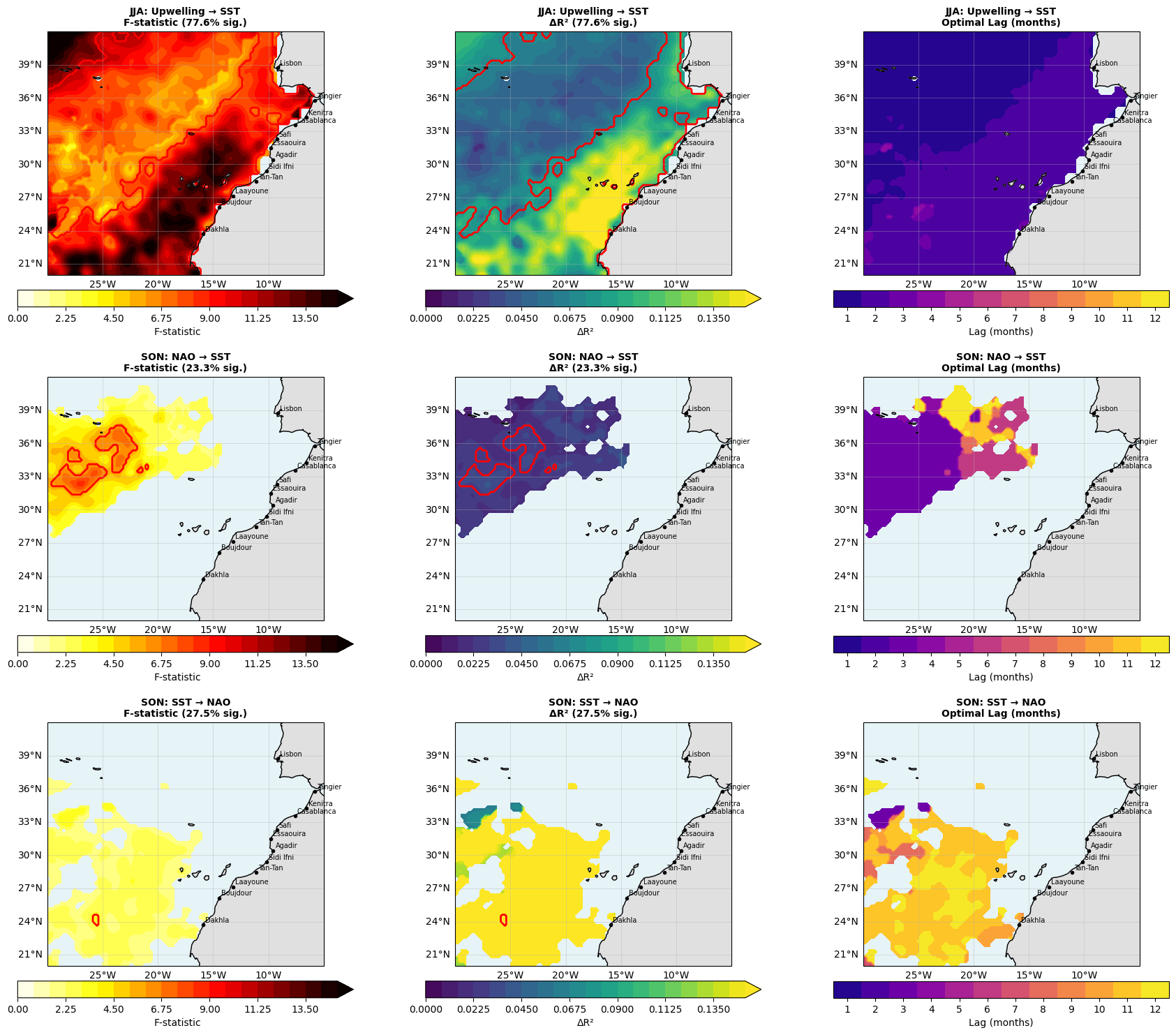}
 \caption{Spatial analysis of Granger causality. \textbf{Top}: CUI $\rightarrow$ SST (JJA); \textbf{Middle}: NAO $\rightarrow$ SST (SON); \textbf{Bottom}: SST $\rightarrow$ NAO (SON). Columns: F-statistic, $\Delta R^2$, optimal lag (months). Red contours: significance.}
  \label{fig:granger_maps}
\end{figure}

\begin{table}[H]
    \centering
    \caption{Significant Granger causality results (\(p < 0.05\))}
    \label{tab:granger}
    \begin{tabular}{lcc}
        \toprule
        \textbf{Relationship} & \textbf{Season} & \textbf{Optimal Lag (months)} \\
        \midrule
        CUI $\rightarrow$ SST & JJA & 1--3 \\
        NAO $\rightarrow$ SST & SON & 2--4 \\
        SST $\rightarrow$ NAO & SON & 7--11 \\
        \bottomrule
    \end{tabular}
\end{table}

\subsubsection{Interpretation of Granger Causality Analysis}

Granger analysis reveals complex directional relationships and distinct time scales for atmosphere-ocean-upwelling coupling.

\paragraph{Direct Local Forcing: CUI $\rightarrow$ SST in Summer (JJA)}
\begin{itemize}
    \item \textbf{Direction and Time Scale:} CUI (Coastal Upwelling Index) causes SST with a \textbf{very short lag (1-3 months)} (Table~\ref{tab:granger}). This short lag (upper right panel of Figure~\ref{fig:granger_maps}) is expected and represents the direct physical mechanism: wind forces upwelling of cold waters, and thermal effect on surface is almost immediate.
    \item \textbf{Location and Intensity:} Causality is strongest (F-statistic and $\Delta R^2 \le 0.13$) along the coast, particularly off Dakhla ($23^\circ$N-$25^\circ$N), where summer upwelling is most intense. The low $\Delta R^2$ suggests that while upwelling causes SST, thermal inertia of surface layer and other atmospheric factors also play an important role in final SST variability.
\end{itemize}

\paragraph{Atmospheric Teleconnection: NAO $\rightarrow$ SST in Autumn (SON)}
\begin{itemize}
    \item \textbf{Direction and Time Scale:} NAO causes SST with a \textbf{lag of 2 to 4 months} (Table~\ref{tab:granger}). This lag (middle right panel of Figure~\ref{fig:granger_maps}) indicates non-immediate atmospheric influence. The NAO signal (which is strong in winter, DJF) propagates and modulates ocean SST, probably via heat flux anomalies or through maintenance of anomalies inherited from previous NAO phase.
    \item \textbf{Location:} Causality is significant (red contours) mainly in the northwest region (toward Canaries and $30^\circ$N-$35^\circ$N), far from the coast. This suggests NAO acts by modulating thermal conditions \textbf{at large scale} in the Canary basin before this signal propagates toward Moroccan coastal waters. The low explanation rate ($\Delta R^2 \approx 0.09$) reinforces the idea of indirect or moderate teleconnection.
\end{itemize}

\paragraph{Ocean Feedback: SST $\rightarrow$ NAO in Autumn (SON)}
\begin{itemize}
    \item \textbf{Direction and Time Scale:} Analysis reveals inverse causality: SST causes NAO with a \textbf{long lag of 7 to 11 months} (Table~\ref{tab:granger}). This long lag (lower right panel of Figure~\ref{fig:granger_maps}) is the signature of significant \textbf{ocean memory}.
    \item \textbf{Feedback Mechanism:} The ocean accumulates (or dissipates) heat over prolonged periods (e.g., SST anomalies of central Atlantic), and this thermal persistence eventually influences atmospheric circulation at basin scale, including configuration of following season's NAO. Causality is strongest in the center (Agadir-Safi region, $28^\circ$N-$32^\circ$N), with $\Delta R^2 \le 0.18$ and high F-statistic, indicating this is one of potential predictability mechanisms of NAO at long lead time.
\end{itemize}

Granger causality analysis highlights the hierarchy of processes: \textbf{rapid local forcing} (CUI $\rightarrow$ SST) dominates in summer, while in autumn, a \textbf{delayed and bidirectional coupling} (NAO $\leftrightarrow$ SST) characterizes the system, with SST playing a crucial memory role for future NAO evolution.

\subsection{Climate Trends and Breakpoint Detection}

\subsubsection{Linear Trends 1978-2024}

Analysis of climate trends reveals contrasting evolution of studied variables (Table~\ref{tab:climate_trends}). Sea surface temperature (SST) shows a significant increasing trend in all seasons, with average increase of \textbf{+0.0736 units/decade}.

\begin{table}[H]
\centering
\caption{Seasonal climate trends 1978-2024 (units/decade)}
\begin{tabular}{lcccc}
\toprule
\textbf{Season}  & \textbf{NAO} & \textbf{Upwelling} & \textbf{SST} \\
\midrule
DJF  & +0.0813 & +0.0918 & +0.0642* \\
MAM  & +0.0064 & -0.0400 & +0.0683* \\
JJA  & -0.0419 & -0.0635* & +0.0774* \\
SON  & +0.0398 & -0.0051 & +0.0846* \\
\midrule
\textbf{Average} & \textbf{+0.0214} & \textbf{-0.0042} & \textbf{+0.0736*} \\
\bottomrule
\end{tabular}
\label{tab:climate_trends}
\end{table}

\noindent*Significant at threshold $p < 0.05$

\textbf{Main observations}:
\begin{itemize}
    \item \textbf{SST}: Generalized and statistically significant increase in all seasons, particularly marked in autumn (+0.0846/decade)
    \item \textbf{Upwelling}: Slight global decreasing trend (-0.0042/decade), significant only in summer (JJA)
    \item \textbf{NAO}: Weak positive global trend (+0.0214/decade), with important seasonal variability
\end{itemize}

\subsubsection{SST Trend Mapping}

Spatial mapping of SST trends (Figure 3.8) reveals \textbf{generalized warming} of the Moroccan Atlantic coast, with average trend of \textbf{+0.149°C/decade}.

\begin{figure}[H]
    \centering
    \includegraphics[width=0.9\textwidth]{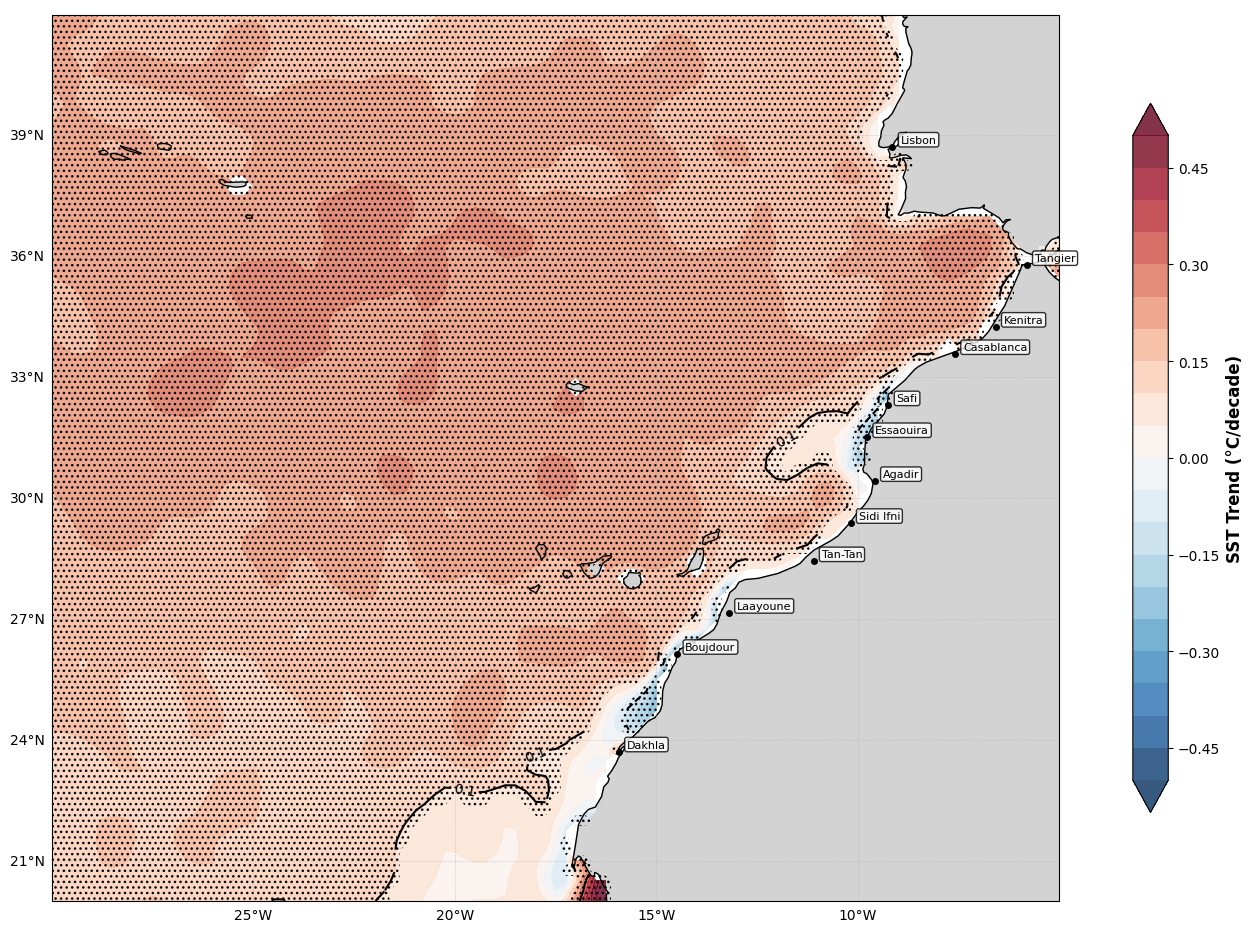}
    \caption{Trends of sea surface temperature (SST) along the Moroccan Atlantic coast (1978-2024). The map shows generalized warming (+0.149°C/decade on average) with 98.2\% of the area warming and only 1.8\% cooling. Hatched areas indicate statistical significance ($p < 0.05$).}
    \label{fig:sst_trends_map}
\end{figure}

\textbf{Spatial characteristics}:
\begin{itemize}
    \item \textbf{Dominant warming}: 98.7\% of study area shows positive trend
    \item \textbf{High significance}: 95.5\% of trends are statistically significant
    \item \textbf{Localized cooling}: Some coastal areas show slight cooling, potentially linked to intensification of upwelling
\end{itemize}

\subsubsection{Breakpoint Detection}

Breakpoint analysis (Figure 3.9) identified several key years in the evolution of climate variables:

\begin{figure}[H]
    \centering
    \includegraphics[width=0.9\textwidth]{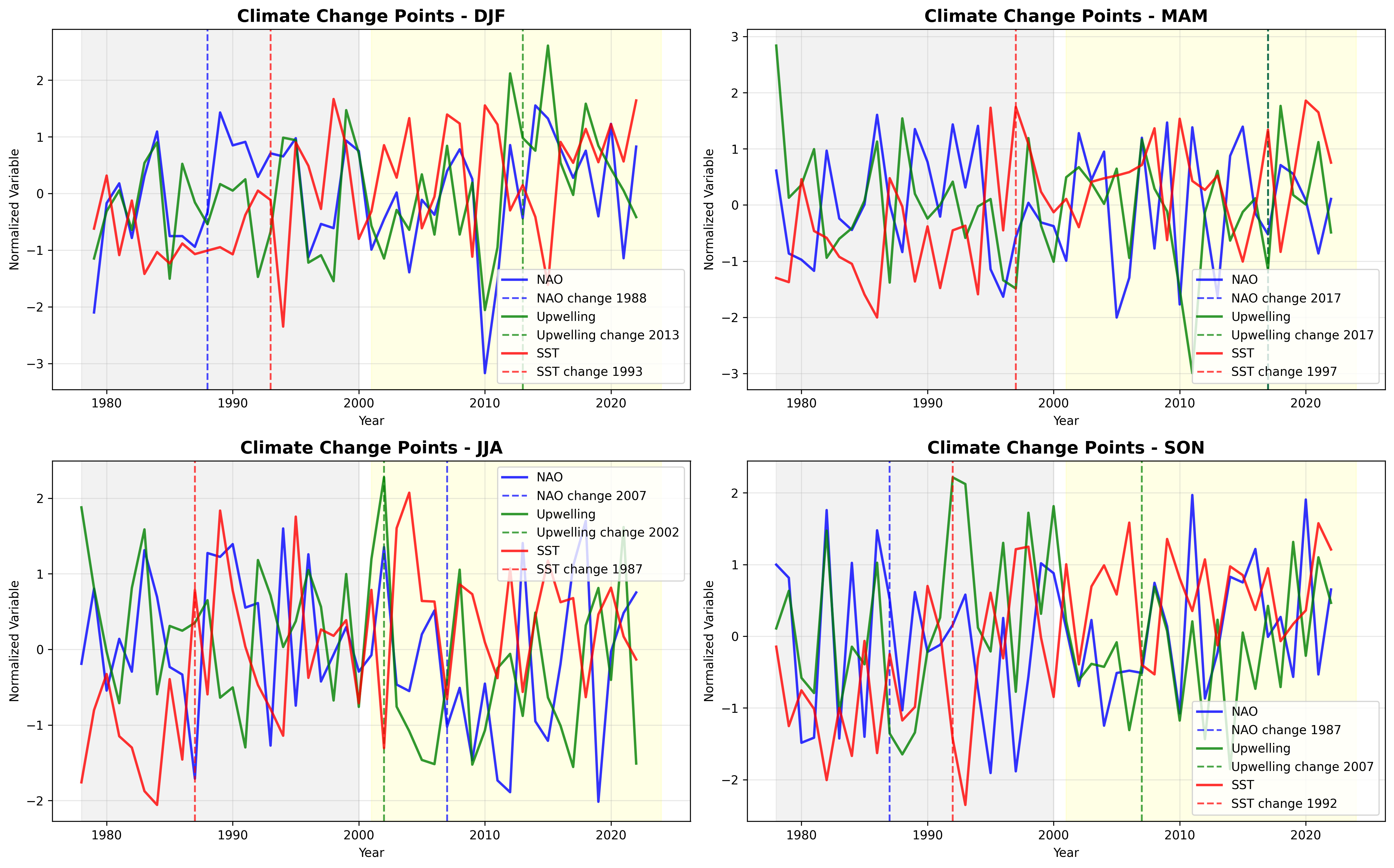}
    \caption{Breakpoints detected in seasonal climate index time series (1978-2024). Dotted vertical lines indicate years of climate regime change detected automatically.}
    \label{fig:change_points}
\end{figure}

\textbf{Significant breakpoints}:
\begin{itemize}
    \item \textbf{SST}: Breakpoints detected around 1992-1997, corresponding to acceleration of regional warming
    \item \textbf{Upwelling}: Regime changes in 2002-2013, potentially linked to modification of wind regimes
    \item \textbf{NAO}: Several breakpoints detected (1987-2017), reflecting decadal variability of the oscillation
\end{itemize}

\subsubsection{Regional Climate Paradox}

Our results highlight a \textbf{climate paradox} in the region:
\begin{itemize}
    \item \textbf{Global warming}: SST shows significant increasing trend (+0.0736 units/decade)
    \item \textbf{Localized intensification of upwelling}: Some coastal areas show cooling despite global warming
    \item \textbf{Confirmation of Bakun hypothesis}: Climate change could reinforce upwelling via intensification of land-sea thermal gradients
\end{itemize}

This analysis confirms that \textbf{climate change significantly modifies ocean-atmosphere dynamics} in the region, with important implications for marine ecosystems and coastal socio-economic activities.

\subsubsection{Mapping of Annual Upwelling Magnitude (1978-2024)}

\begin{figure}[H]
    \centering
    \includegraphics[width=0.9\textwidth]{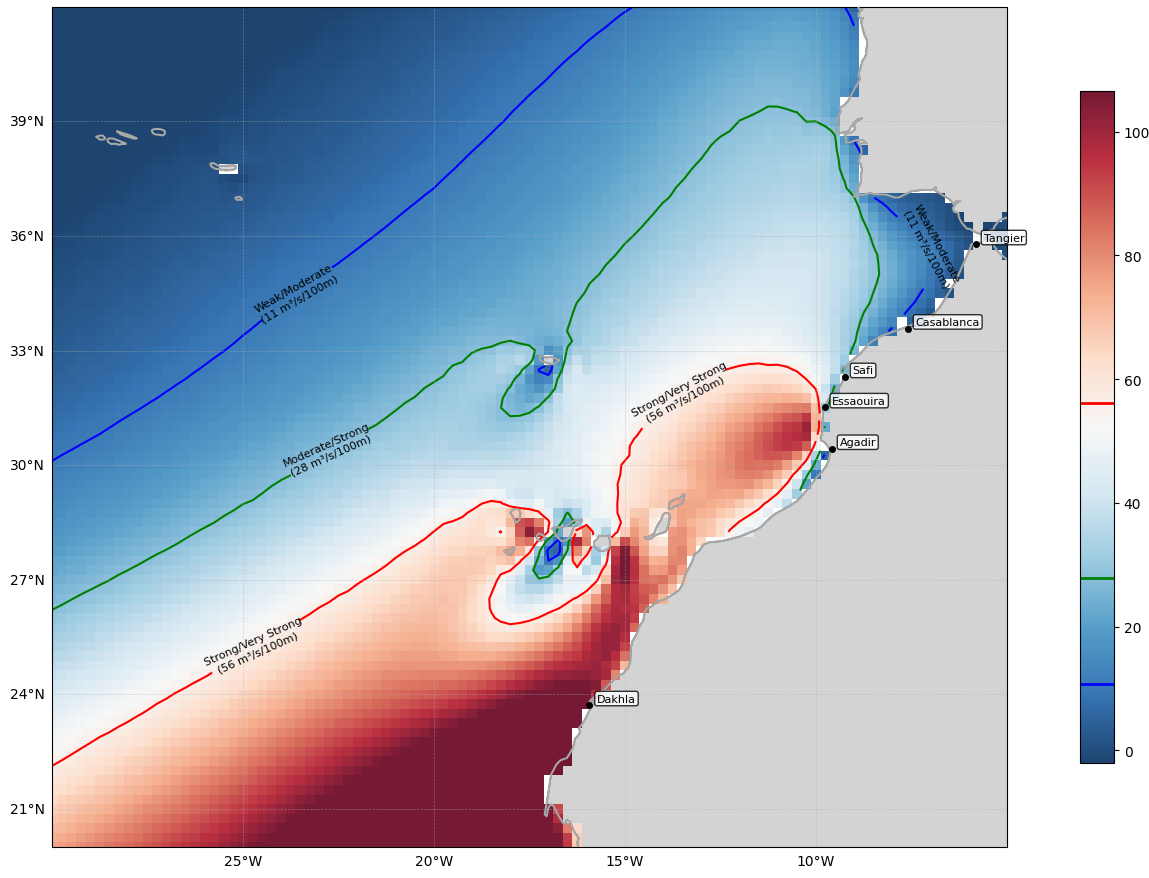}
    \caption{Annual mean magnitude of upwelling transport (Ekman Transport) along the Moroccan Atlantic coast (1978-2024). Color indicates intensity, and curves (blue, green, red) represent dynamic classification thresholds based on percentiles of calculated data (P25, P50, P75).}
    \label{fig:upwelling_magnitude}
\end{figure}

Upwelling intensity was calculated over the period 1978-2024 using the Ekman transport formula ($Q_E$). Figure~\ref{fig:upwelling_magnitude} presents the annual mean magnitude of this transport, expressed in cubic meters per second per 100 meters of coast ($\text{m}^3/\text{s}/100\text{m}$).

\paragraph{Interpretation of Magnitude and Upwelling Zones}

Spatial analysis of Figure~\ref{fig:upwelling_magnitude} and obtained descriptive statistics (\textnormal{Mean: $37.1 \pm 34.5$ $\text{m}^3/\text{s}/100\text{m}$}) highlight key zones of maximum intensity and geographic distribution of the phenomenon:

\begin{itemize}
    \item \textbf{Extremely Strong Upwelling Zones (South)}: The strongest intensity is clearly localized in southern Morocco, particularly off Dakhla and regions extending southwest. It is in this region that Ekman transport exceeds the threshold of \textbf{56 $\text{m}^3/\text{s}/100\text{m}$} (red curve), corresponding to the 75th percentile (\textit{Very Strong Upwelling}). This is directly linked to ideal coastline configuration and constancy and strength of Trade Winds in this region.
    \item \textbf{Moderate to Strong Zones (Center)}: Along central coasts, between Agadir and Safi, upwelling presents moderate to strong intensity, mainly situated between thresholds of \textbf{27.9} and \textbf{56.2 $\text{m}^3/\text{s}/100\text{m}$} (between green and red curves). Upwelling here is seasonal, heavily dependent on summer position of Azores High.
    \item \textbf{Weak to Moderate Zones (North)}: North of Safi, and particularly near Casablanca and Tangier, upwelling magnitude decreases. Transport is often below the threshold of \textbf{10.7 $\text{m}^3/\text{s}/100\text{m}$} (blue curve, \textit{Weak Upwelling}), or even negative, due to modification of coastline orientation and increased wind variability. Low transport in this zone explains warmer SST and weak signature of upwelling in SST data at these latitudes.
\end{itemize}

Thresholds calculated dynamically from actual data (P25: $10.7$, P50: $27.9$, P75: $56.2$ $\text{m}^3/\text{s}/100\text{m}$) allow precise delimitation of ecoregions where upwelling effects (i.e., SST cooling and nutrient supply) are most pronounced. Dominance of strong signal in the South confirms this is the main coastal production zone of the Moroccan marine ecosystem.

\subsection{Discussion and Comparison with Literature}

\subsubsection{Comparison with \textcite{wang2004}}

\textcite{wang2004} demonstrate, via Granger causality, that SST anomalies in the North Atlantic contain predictive information for winter NAO with lags of 6-12 months, particularly in the Gulf Stream region. Our results validate and \textbf{extend} this feedback:

\begin{itemize}
    \item \textbf{Agreement}: SST $\rightarrow$ NAO causality significant in SON with lag 7-11 months (Figure~\ref{fig:granger_maps}, bottom), consistent with their ocean memory, centered on Gulf Stream but localized off Morocco (25°-30°N).
    \item \textbf{Novelty}: Seasonally shifted signal (autumn $\rightarrow$ following winter), and inclusion of CUI showing rapid local forcing.
    \item \textbf{Interpretation}: Moroccan SST, cooled by upwelling, can modulate air-sea thermal gradients and influence NAO via Ferrel cells.
\end{itemize}

> \textit{« [...] preceding SST anomalies have a statistically significant causal effect on the wintertime NAO. However, the causal relation [...] is limited to the Gulf Stream extension »} \parencite[p.~4752]{wang2004}.

\subsubsection{Comparison with \textcite{narayan2010}}

\textcite{narayan2010} identify a \textbf{trend toward upwelling intensification} off NW Africa (1960-2001), but an \textbf{ambiguous relationship with NAO} (negative wind-NAO correlation, absence of SST-NAO link).

\begin{itemize}
    \item \textbf{Agreement}: Our JJA maps (Figure~\ref{fig:cui_corr}a) show strong SST-CUI coupling (\(r = -0.46\)), confirming local intensification. Weak NAO-CUI correlation in JJA/MAM supports their ambiguity.
    \item \textbf{Validation}: Absence of CUI $\sim$ NAO maps due to weak correlations, aligned with their conclusion: upwelling is driven by local winds, not by NAO.
    \item \textbf{Extension}: Our SST $\rightarrow$ NAO causality (lag 7-11 months) adds ocean feedback, potentially linked to Bakun (1990) hypothesis on warming $\rightarrow$ land-sea gradient $\rightarrow$ strengthened winds.
\end{itemize}

> \textit{« The relationship of the North Atlantic Oscillation (NAO) with coastal upwelling off NW Africa turned out to be ambiguous [...] »} \parencite[p.~815]{narayan2010}.

\subsubsection{Comparison with \textcite{hilmi2021}}

\textcite{hilmi2021} document \textbf{quasi-permanent upwelling activity in the south (Dakhla)} with summer peak (1967-2019).

\begin{itemize}
    \item \textbf{Complete agreement}: Our JJA maps (Figure~\ref{fig:cui_corr}a) show \(r = -0.46\) off Dakhla-Safi, and CUI $\rightarrow$ SST causality at 1-3 months. Interannual variability with strong periods: 1967-1980, 2009-2019.
    \item \textbf{Spatial extension}: Significant signal up to 35°N in winter (DJF, \(r = -0.41\)), and clear north-south gradient in new maps.
\end{itemize}

\subsection{Synthesis of Causal Relationships}

\begin{enumerate}
    \item \textbf{Rapid local forcing}: CUI $\rightarrow$ SST (JJA/SON, lag 1-3 months) — summer upwelling cools coast, aligned with \textcite{belmajdoub2023}.
    \item \textbf{Seasonal teleconnection}: NAO $\rightarrow$ CUI/SST (DJF, \(r = +0.65/-0.37\)) — reinforces winter trade winds.
    \item \textbf{Interannual feedback}: SST $\rightarrow$ NAO (SON, lag 7-11 months) — ocean memory modulating NAO, extending \textcite{wang2004}.
\end{enumerate}

\subsection{Socio-economic Implications}

Spatial and seasonal structures highlighted in this work translate to differentiated effects on fishing and aquaculture activities along the Moroccan Atlantic coast. The north-center-south gradient identified by mean upwelling magnitude (Figure~\ref{fig:upwelling_magnitude}), combined with observed climate trends over the period 1978-2024, directly conditions usage distribution and vulnerability of economic sectors dependent on the ocean.

\paragraph{Artisanal fishing (barques)}
Artisanal fishing, concentrated mostly in the coastal band, is directly exposed to local variations in marine productivity. Southern and central-southern zones, characterized by strong to very strong upwelling and rapid CUI $\rightarrow$ SST causality (lag 1-3 months in JJA), constitute the main poles of artisanal production.  
However, results show that \textbf{significant decrease in summer upwelling} observed in JJA, combined with \textbf{generalized increase in SST}, is likely to affect seasonal availability of coastal pelagic resources. This situation increases vulnerability of small artisanal units, whose activity heavily depends on stability of local environmental conditions.

\paragraph{Industrial fishing}
Industrial fishing operates at larger spatial scales and benefits from greater adaptation capacity to stock displacements. Southern zones, where upwelling is quasi-permanent and variability is dominated by local forcing rather than NAO, offer relative stability of fishing habitats.  
Nevertheless, analysis highlights a \textbf{winter NAO $\rightarrow$ SST teleconnection} with delayed effect, suggesting that thermally generated anomalies at large scale can influence distribution of stocks exploited by the industrial fleet at interannual scale. This implies increased need for climate forecasting tools integrated into industrial fishing campaign planning.

\paragraph{Coastal aquaculture}
Results indicate that aquaculture is particularly sensitive to \textbf{SST warming} and persistent thermal anomalies, especially in northern and central zones where upwelling is weak or intermittent.  
For example, farming of \textbf{sole}, a benthic species with limited thermal tolerance, can be affected by warming episodes observed in SON and by reduction of summer cooling linked to weakening of upwelling. Lags identified between CUI and SST imply that unfavorable conditions can persist several months after an atmospheric anomaly, complicating management of aquaculture production cycles.

\paragraph{Integrated reading of results}
The set of results suggests that:
\begin{itemize}
    \item the \textbf{southern coast} retains a relative productive advantage thanks to dominance of local upwelling forcing;
    \item the \textbf{center} appears as a transition zone, sensitive to seasonal variations of CUI;
    \item the \textbf{north} is more exposed to thermal anomalies and climate teleconnections, increasing vulnerability of coastal activities.
\end{itemize}

These elements highlight the interest of \textbf{differentiated and spatially targeted management} of marine resources, based on climate indicators identified in this work (CUI, SST, seasonal lags), to anticipate impacts of climate change on coastal socio-economic activities.

\chapter{General Conclusion and Perspectives}

\section{Synthesis of Main Results}

This study analyzed interactions between the North Atlantic Oscillation (NAO), coastal upwelling index (CUI), and sea surface temperature (SST) along the Moroccan Atlantic coast over the period 1978-2024. The main conclusions are:

\subsection{Seasonal variability of ocean-atmosphere relationships}
\begin{itemize}
    \item \textbf{Winter (DJF)}: Dominant NAO-CUI teleconnection (\(r = +0.65\)), reinforcing trade winds and inducing significant cooling (\(r = -0.37\) NAO-SST; \(r = -0.41\) CUI-SST).
    \item \textbf{Spring (MAM)}: Persistent NAO influence on SST (\(r = -0.40\)), but decoupling with CUI (\(r = +0.07\)).
    \item \textbf{Summer (JJA)}: Peak local upwelling (\(r = -0.46\) CUI-SST), predominant coastal dynamics.
    \item \textbf{Autumn (SON)}: Cumulative effect of upwelling (\(r = -0.57\) CUI-SST), moderate re-emergence of NAO (\(r = +0.32\)).
\end{itemize}

\subsection{North-south spatial gradient}
The eight spatial correlation maps reveal a clear latitudinal gradient: quasi-permanent upwelling in south (Dakhla), maximum summer coupling in center (Safi-Essaouira), more marked NAO teleconnection in north (Casablanca-Tangier).

\subsection{Climate trends 1978-2024}
\begin{itemize}
    \item \textbf{Significant SST warming}: +0.0736 units/decade on average, with acceleration since the 1990s.
    \item \textbf{Summer decrease in upwelling}: -0.0635 units/decade in JJA (\(p < 0.05\)).
    \item \textbf{Weak positive NAO trend}: +0.0214 units/decade (non-significant).
\end{itemize}

\subsection{Upwelling mapping}
Annual upwelling magnitude (1978-2024) shows average intensity of \(37.1 \pm 34.5\ \text{m}^3/\text{s}/100\text{m}\) with classification thresholds based on percentiles:
\begin{itemize}
    \item P25 (10.7): weak threshold
    \item P50 (27.9): moderate threshold
    \item P75 (56.2): strong threshold
\end{itemize}

\subsection{Granger causality analysis}
\begin{enumerate}
    \item \textbf{Rapid local forcing}: CUI \(\rightarrow\) SST (lag 1-3 months, JJA/SON) — immediate adiabatic response.
    \item \textbf{Seasonal teleconnection}: NAO \(\rightarrow\) CUI/SST (DJF) — winter wind reinforcement.
    \item \textbf{Interannual feedback}: SST \(\rightarrow\) NAO (lag 7-11 months, SON) — ocean memory influencing following NAO.
\end{enumerate}

\section{Validation with Literature}

These results validate and extend previous work:
\begin{itemize}
    \item \textcite{wang2004}: Confirmation of SST-NAO feedback with lag 7-11 months.
    \item \textcite{narayan2010}: Validation of local upwelling intensification.
    \item \textcite{hilmi2021}: Confirmation of quasi-permanent activity in the south.
    \item \textcite{belmajdoub2023}: Extension with new EBU index.
\end{itemize}

\section{Environmental and Socio-economic Implications}

\subsection{Impacts on fishing}
\begin{itemize}
    \item \textbf{Artisanal fishing} ($\approx$ 12,000 barques): Strong dependence on intense upwelling zones (South/Center), ensuring 70\% of sardine catches.
    \item \textbf{Industrial fishing} ($\approx$ 455 vessels): Operation in persistent upwelling zones, national economic pillar.
    \item \textbf{Predictability}: Integration of CUI (lag 3 months) and SST (lag 9 months) for activity optimization.
\end{itemize}

\subsection{Regional climate paradox}
\begin{itemize}
    \item \textbf{Global warming}: +0.149°C/decade on average (98.2\% of the area).
    \item \textbf{Localized cooling}: In intense upwelling zones, confirming Bakun (1990) hypothesis.
    \item \textbf{Implications}: Potential ecosystem modification and need for spatially differentiated approaches.
\end{itemize}

\subsection{Other sectors}
\begin{itemize}
    \item \textbf{Coastal tourism}: Persistent cold waters (<20°C in summer) potentially affecting beach attractiveness.
    \item \textbf{Aquaculture}: Risks of thermal stress related to SST-NAO variability.
    \item \textbf{Wind energy}: Opportunities in zones of strengthened winds (Tangier-Safi).
\end{itemize}

\section{Study Limitations}

\begin{itemize}
    \item \textbf{Spatial resolution}: ERA5 grid (\SI{0.25}{\degree}) may smooth fine coastal phenomena (<10 km).
    \item \textbf{Analysis period}: Extended to 1978-2024, but absence of future projections (CMIP6).
    \item \textbf{Biological variables}: No direct integration of chlorophyll or actual catches.
    \item \textbf{Statistical causality}: Inherent limitations of Granger test for establishing direct physical relationships.
\end{itemize}

\section{Research Perspectives}

\begin{enumerate}
    \item \textbf{Coupled modeling}: Use ROMS-WRF to simulate RCP/SSP scenarios and assess future upwelling evolution.
    \item \textbf{Biological integration}: Correlate CUI/SST with phytoplankton biomass (MODIS/OCM satellite) and INRH landings.
    \item \textbf{Operational forecasting}: Develop seasonal bulletin integrating CUI (lag 3 months) and SST (lag 9 months) for INRH and fishermen.
    \item \textbf{In situ studies}: Targeted oceanographic campaigns (buoys, gliders) to validate high-resolution indices.
    \item \textbf{Multi-scale analysis}: Integrate other climate indices (AMO, EAWR) for more comprehensive understanding.
\end{enumerate}

\section{Final Conclusion}

This study demonstrates that \textbf{climate change significantly modifies interactions between NAO, upwelling, and SST} along the Moroccan Atlantic coast. The observed climate paradox—global warming coexisting with localized cooling in upwelling zones—highlights the complexity of regional climate change impacts.

The region constitutes a \textbf{natural laboratory} for studying ocean-atmosphere couplings under climate change effects. Obtained results provide a solid scientific basis for adaptive management of marine resources, founded on seasonal forecasting and modeling, essential for sustainable exploitation facing future climate changes.

Validation of identified causal relationships and their integration into operational tools represent the next crucial steps for transforming this research into concrete benefits for economic sectors dependent on coastal marine ecosystems.

\end{document}